\documentclass[twocolumn,times,tighten]{aastex63}

\usepackage{color}
\usepackage{natbib}
\usepackage{graphicx}

\shorttitle{Interstellar Medium toward 2MASS $J$17470898$-$2829561}
\shortauthors{T. R. Geballe et al.}
\begin{document}
\title{The Interstellar Medium toward the Galactic Center Source 2MASS J17470898$-$2829561}
\author{T. R. Geballe}
\affiliation{Gemini Observatory/NSF's NOIRLab, 670 N. A'ohoku Place, Hilo, Hawai'i, 96720, USA}
\email{tom.geballe@noirlab.edu}
\author{Yvonne Pendleton}
\affiliation{NASA Ames Research Center, Moffett Field, CA 94035, USA}
\author{Jean Chiar}
\affiliation{Diablo Valley College, 321 Golf Club Rd, Pleasant Hill, CA 94523, USA}
\author{Alexander G. G. M. Tielens}
\affiliation{Leiden University, Niels Bohrweg 2, 2333 CA Leiden, The Netherlands}

\begin{abstract}
We describe and discuss remarkable infrared spectra, covering key portions of the $2-5$~$\mu$m  wavelength interval, of the probable OH/IR supergiant 2MASS $J$17470898$-$2829561 (2M1747), located in direction of the Sgr B molecular cloud complex within the Central Molecular Zone (CMZ) of the Galaxy.  This star was originally singled out for examination based on its suitability for spectroscopy of lines of H$_3^+$ in the CMZ. Analysis of the spectra shows that 2M1747 is deeply embedded within Sgr B1, with $A_V$  $\gtrsim$ 100 mag, making it the only star within Sgr B for which infrared spectra have been obtained at present, and thereby a unique infrared probe of the dense interstellar medium within the CMZ. Despite the high extinction, spectra of 2M1747 reveal a veiled photosphere in the $K$ band and circumstellar gas in the $M$ band, giving clues as to its nature.  Its $3.5-4.0~\mu$m spectrum contains the strongest absorption lines of  H$_3^+$ observed toward any object to date. The $4.5-4.8$ $\mu$m spectrum has impressively deep and wide absorption lines of interstellar CO, most of which arise in dense gas within Sgr B1. The $3-5$ $\mu$m spectrum also contains several solid-state absorption features, which are characteristic of both dense and diffuse clouds, and which raise questions about the identifications of some of these features. We discuss the nature of the star, the extinction to it, the extinction law for dust in the CMZ, and the identifications of the various solid-state features and where they are produced along this complex line of sight.

\end{abstract}

\keywords{astrochemistry ---  Galaxy: center --- infrared: stars --- ISM: lines and bands --- ISM:
molecules}

\section{Introduction}

The Central Molecular Zone (CMZ) of the Galaxy \citep{mor96} is a region of diameter $\sim$300 pc and thickness
$\sim$50 pc centered on the supermassive black hole, Sgr A*, at the Galactic center (GC), here assumed to be at a distance of 8 kpc. \cite{geb99} discovered unusually high column densities of H$_3^+$ toward two stars in the CMZ, one located in the Central Cluster of hot luminous stars immediately surrounding Sgr A*, the other in the Quintuplet Cluster, 30 pc distant from Sgr A* on the plane of the sky. Using high-resolution infrared spectroscopy of three key absorption lines of H$_3^+$ at 3.5337, 3.6205, and 3.7155 $\mu$m in its fundamental ($\Delta v$ = 1) vibration-rotation band together with spectroscopy of absorption lines from low rotational levels of the first overtone ($\Delta v$ = 2) band of CO  near 2.34 $\mu$m,  on sightlines towards several hot stars and dust-embedded stars within $\sim$30 pc of Sgr A*, \cite{oka05} and \cite{got08} established that the absorbing  H$_3^+$ in the CMZ on sightlines toward this region is located in warm (T$\sim$200 K) and diffuse ($n \lesssim 10^{2}$ cm$^{-3}$) interstellar gas within the CMZ. Based on the observed large column densities of the H$_3^+$, the lengths of the absorbing columns appeared to be large fractions of the radius of the CMZ. Although information was lacking on sightlines spread across the full longitudinal extent of CMZ, that finding suggested that warm diffuse gas fills a large fraction of the volume of the CMZ. 

In order to test whether indeed the warm diffuse gas fills much of the CMZ, it was necessary to find, within the CMZ but much more widely distributed in Galactic longitude, bright stars whose intrinsic spectra are simple enough for them to be used to accurately measure the strengths and profiles of the above three absorption lines of H$_3^+$. \cite{geb19} completed a low-resolution $K$-band spectroscopic survey of ~500 candidate stars in the CMZ with $L$ magnitudes brighter than 7.7, and discovered ~30 such stars, scattered across the CMZ. Almost all of these stars were previously unknown except for their designations and $JHK$ photometry in the catalog of the Two-Micron All Sky Survey \cite[2MASS,][]{skr06} and longer-wavelength photometry of them obtained by the Spitzer Space Telescope as part of the Galactic Legacy Infrared Midplane Survey Extraordinaire \cite[GLIMPSE,][]{ram08}.

Many of the newly discovered stars are emission-line stars of various types. Some of them are embedded in optically thick and warm circumstellar shells; many of that subset are dusty Wolf-Rayet stars. The remainder of the newly discovered stars are red giants with warm circumstellar shells that heavily veil their complex photospheric $K$-band absorption spectra. For each of these, high-resolution spectroscopy of the above interstellar lines of H$_3^+$ produces spectra that are relatively straightforward to analyze because they are uncontaminated by photospheric lines. The spectra of the H$_3^+$ lines and CO first overtone band lines toward most of these stars revealed that the H$_3^+$ in front of them is also situated in long columns of warm and diffuse interstellar gas, similar to those seen earlier toward stars nearer the center of the CMZ. This finding has demonstrated the pervasiveness of a warm and relatively low-density interstellar environment in the CMZ, estimated by  \cite{oka19} to have a filling factor of $\sim$ 2/3. 

Comparison of H$_3^+$ and CO velocity profiles toward almost all of these objects indicates little or no absorption by CO in dense molecular gas within the CMZ \citep{oka19}, with one of the few exceptions being the subject of this paper. The lack of absorption by overtone band lines of CO in dense clouds could be partly a selection effect. A number of giant molecular clouds (GMCs) are located within the CMZ; the dust they contain produces significant extinction in addition to that produced by dust in the foreground spiral arms and in the diffuse interstellar medium (ISM) of the CMZ. Thus, their effect is to significantly reduce the apparent brightnesses of stars within or behind them. However, \citet[][see their Section 1.3 and references therein]{oka19} concluded that the filling factor of  clouds with $n > 3 \times 10^{3}$ cm$^{-3}$ in the CMZ is at most a few percent, too small to require that their conclusion regarding the large filling factor of the warm diffuse gas in the CMZ be modified. Thus while the name CMZ originally was bestowed on the region because of the large mass of molecular gas in dense clouds there, it is actually diffuse molecular gas, which has a total mass roughly comparable to that of the gas in dense molecular clouds, which fills most of the volume.

The reader will note that in discriminating between dense ($n \gtrsim 5 \times 10^{2}$ cm$^{-3}$) and diffuse 
 ($n \lesssim 5 \times 10^{2}$ cm$^{-3}$) molecular gas \citep[see Fig. 15 of][]{mil20}, in addition to the above H$_3^+$ lines the $v$ = $2-0$ band of CO (centered near 2.35 $\mu$m) has been employed rather than the $v$ = $1-0$ band (centered near 4.7 $\mu$m). The square of the transition dipole moment for the CO overtone band is $\sim$260 times less than that of the fundamental band, so that the overtone band strength is $\sim$130 times less than the fundamental \citep{zou02}. Due in part to the weakness of the CO overtone, the CO fundamental band is generally used for infrared studies of CO in circumstellar and interstellar environments. However, because of the high extinction toward sources in the GC even the low abundance of CO in diffuse molecular gas, where C is mostly atomic, can produce deep absorptions in fundamental band lines, which makes it difficult to use lines of that band for discriminating between CO in dense molecular gas and CO in diffuse molecular gas and to determine column densities. On the other hand, absorption in the low-$J$ lines of the overtone band by CO in diffuse gas are very weak and often not detected, while the same absorption lines viewed through clouds of dense gas, although optically thin, are sufficiently strong that they can be used to obtain accurate column densities.
 
\begin{figure*}[]
\begin{center}
\includegraphics[angle=0,width=0.65\textwidth]{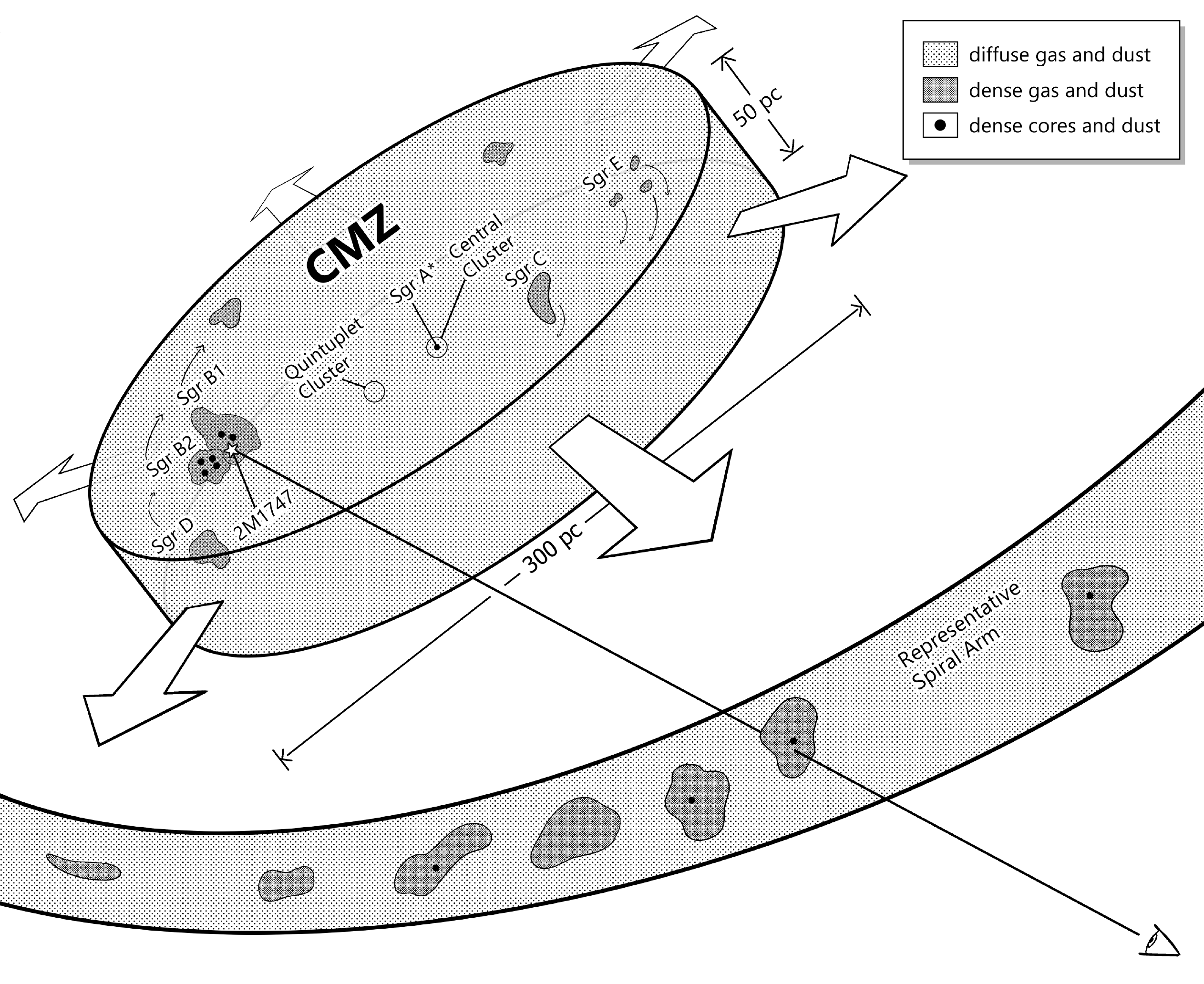}
\end{center}
\caption{Simplified diagram of the disk-shaped CMZ and the sightline toward 2M1747 showing locations of potentially absorbing gas and dust. Diffuse molecular gas in the CMZ is denoted by light shading, several giant molecular clouds in it by medium shading, and representative dense cores by small dark circles.  A single representative foreground spiral arm containing all of these components is also shown, with the same shadings to indicate diffuse and dense molecular gas and dense cores. The low-density interarm medium is depicted with no shading. The line of sight to 2M1747 passes through all of these components except dense cores. Galactic center radio sources Sgr B - E and their associated  giant molecular clouds are labeled, as are the Central and Quintuplet clusters referred to in this paper.  Other known giant molecular clouds in the CMZ are not shown. Typical gas temperatures throughout most of the volume of the CMZ and in diffuse and dense molecular gas in the spiral arms are given in Section 5.3. Orbital motions of the dense clouds in the CMZ and the radial expansion of the diffuse gas in the CMZ are denoted by arrows.}
\label{lpspec}
\end{figure*}

\section{2MASS $J$17470898$-$2829561}

By far the most glaring exception to there being little or no absorption by H$_3^+$ fundamental and CO overtone lines in dense CMZ gas toward the stars identified in \cite{geb19} is the star 2MASS $J$17470898$-$2829561, hereafter 2M1747, but referred to as Star $\iota$ in \cite{oka19} and \cite{oka20}. Toward this star absorption by the strongest H$_3^+$ lines and the low-$J$ CO overtone lines extend continuously over a velocity range of over 200 km s$^{-1}$. The  profiles of  both contain a multitude of  velocity components \cite{geb10}. These interstellar CO and H$_3^+$ lines have much larger absorption equivalent widths than observed to date in the ISM toward any star in the Galactic center or elsewhere.

The column density of CO toward 2M1747 derived from the overtone band lines is huge, $2.7 \times 10^{19}$ cm$^{-2}$ \citep{oka19}. In view of the strengths of the overtone lines, almost all of this CO must reside in dense clouds, as argued at the end of the previous section. This can also be concluded based on extinction arguments. The ratio of dust extinction to CO column density in dense clouds is much smaller than in diffuse clouds, where very little carbon is in the form of CO. If much of the CO observed toward 2M1747 were in diffuse clouds the extinction to the star would be far too large for it to be detectable in the $K$ band. See Section 5.1 for further discussion.

In the plane of the sky 2M1747 is located 85 pc from of Sgr A* in the direction of the Sgr B molecular cloud complex. This region, with linear dimensions of $\sim$30 pc, is dominated by two giant molecular clouds of roughly equal size,  Sgr B1 and Sgr B2.  Based on the high angular resolution radio continuum image of \citet[][see their~Fig. 12]{lan10}, 2M1747 is located within Sgr B1, but close  to its northern edge. Both Sgr B1 and Sgr B2 are sites of ongoing star formation. Although this activity is considerably more intense in Sgr B2,  the presence of compact regions of ionized gas  and H$_2$O maser emission \citep{meh93} within Sgr B1 implies that star formation there is still ongoing \cite[see][]{bar17}.

Unsurprisingly 2M1747, like all of the other suitable stars found by \cite{geb19}, has no optical counterpart. Infrared photometry of 2M1747 by 2MASS and GLIMPSE resulted in the following values: $J$ = 17.18, $H$ = 15.35, $K_{S}$ = 10.45, IRAC1 = 6.58, IRAC2 = 5.79, IRAC3 = 3.28 (the IRAC filters are centered at 3.6 $\mu$m, 4.5 $\mu$m, and 5.7 $\mu$m, respectively). The 2MASS $J$-band image reveals a crowded field surrounding the coordinates of 2M1747 and does not show a discernible pointlike object at its location; thus the above $J$ magnitude is highly suspect. 2M1747 is present in the 2MASS $H$-band image, but also is in a crowded field with many nearby stars of brightness similar to 2M1747, suggesting that its listed $H$ magnitude is also affected by contamination.  Only in the $K_{S}$-band image does 2M1747 stand out clearly in the 2MASS and is photometry of it reliable. 

The cause of the very red infrared colors of 2M1747 is a combination of the star's low photospheric temperature, high extinction, and thermal emission from its warm circumstellar dust. Column densities of gas through the cloud cores in Sgr B2 have been reported to be greater than $\sim$1 $\times 10^{25}$ cm$^{-2}$ \citep{qin11}, which implies visual extinctions $\gtrsim$10$^{4}$ mag assuming the standard Galactic gas-to-dust ratio \citep{boh78}, and even higher if the metallicity in the Galactic center is supersolar, as is suspected \citep[see, e.g. Section A.2 of][]{oka19}. Sgr B1 is believed to be more evolved and its gas more dispersed than Sgr B2. Nevertheless, the far-infrared mapping of \citet[][their figure axes are in 1950 coordinates]{lis01} indicates that the extinction through the region of Sgr B1 where 2M1747  is located (1950 coordinates 17:43:59.01,  $-$28:28:53.1), corresponding in their Figs. $1-3$ to locations near the top left of the frames containing positions $e$ and $f$, is only an order of magnitude less, and thus is still formidable.  The nearest compact and dense clumps of  gas found in the 1.1 mm continuum emission survey of \cite{bal10}, their Clumps 190 and 193, with estimated visual extinctions exceeding 10$^{3}$ mag, are 3 arcminutes distant. Even though 2M1747 is not within or behind a dense molecular core, clearly it cannot be located behind or even near the center of the more extended Sgr B1 dense gas observed by \cite{lis01} or it would not be so bright in the $K$ band.  Thus although deeply embedded in Sgr B1, it must reside in front of most of the dust associated with Sgr B1 on that sightline, but not behind any dense cores in Sgr B1.

Figure 1 is a simplified view of the line of sight toward 2M1747 showing the various interstellar environments that might contribute absorption by molecules in the gas phase and by solid-state matter, including both the dust grains themeselves and ices on them.  In the figure the CMZ is depicted as described by \cite{oka19}, with warm diffuse gas of mean density $\sim$50 cm$^{-3}$ expanding radially from a location or region near the center of the CMZ and taking up most of the volume of the CMZ. Several of the best known dense cloud complexes within the CMZ are shown, as are representative dense cores within them. A representative foreground spiral arm is also shown, containing diffuse and dense clouds at lower temperatures than those in the CMZ.  

The coordinates of 2M1747 are coincident to within a fraction of an arcsecond with OH 0.548-0.059, an OH/IR star \citep{lin92}. In addition to the OH maser emission, \cite{shi97} found SiO maser emission in this source, which \cite{deg97} identified with an infrared point source whose $H$ and $K$ magnitudes are roughly those listed  by 2MASS (albeit a factor of 2 brighter at $H$). \cite{sjo02} detected H$_{2}$O maser emission at a location 5 $\pm 5$ arcseconds distant and concluded it was associated with the OH/IR star. All of the above authors identified 2M1747 as a late-type star. Some of those authors have suggested, based on its infrared brightness and intense SiO maser emission, that this star is in the foreground, well outside of the Galactic center. The data presented here and by \citet{geb10}, \cite{got11}, and \cite{oka19} make it clear that this is not the case.  

This paper presents infrared spectra of 2M1747 in several spectral segments within the $2-5~\mu$m wavelength range.  The spectra were obtained at three telescopes and at a variety of resolutions. Section 3 is a review of previously published spectroscopy of 2M1747 and what it reveals about the nature of the star and about the properties of the interstellar gas that produces absorption in lines of H$_3^+$ and CO.  Section 4 presents and describes more recently acquired spectra in the following order: the $K$ band, which pertains mainly to the nature of the star;  the $M$ band, which contain features due to both interstellar gas and dust; and the $L$ band, which is dominated by absorption features due to dust. Section 5 contains rough calculations of the extinction to 2M1747 and the luminosity of the star, and how these derived values interact with one another, together with discussions of the likely locations on the line of sight to 2M1747 of the dust grains that produce several of the solid-state absorption features in the $L$ and $M$ bands.

\begin{figure}[]
\includegraphics[angle=0,width=0.47\textwidth]{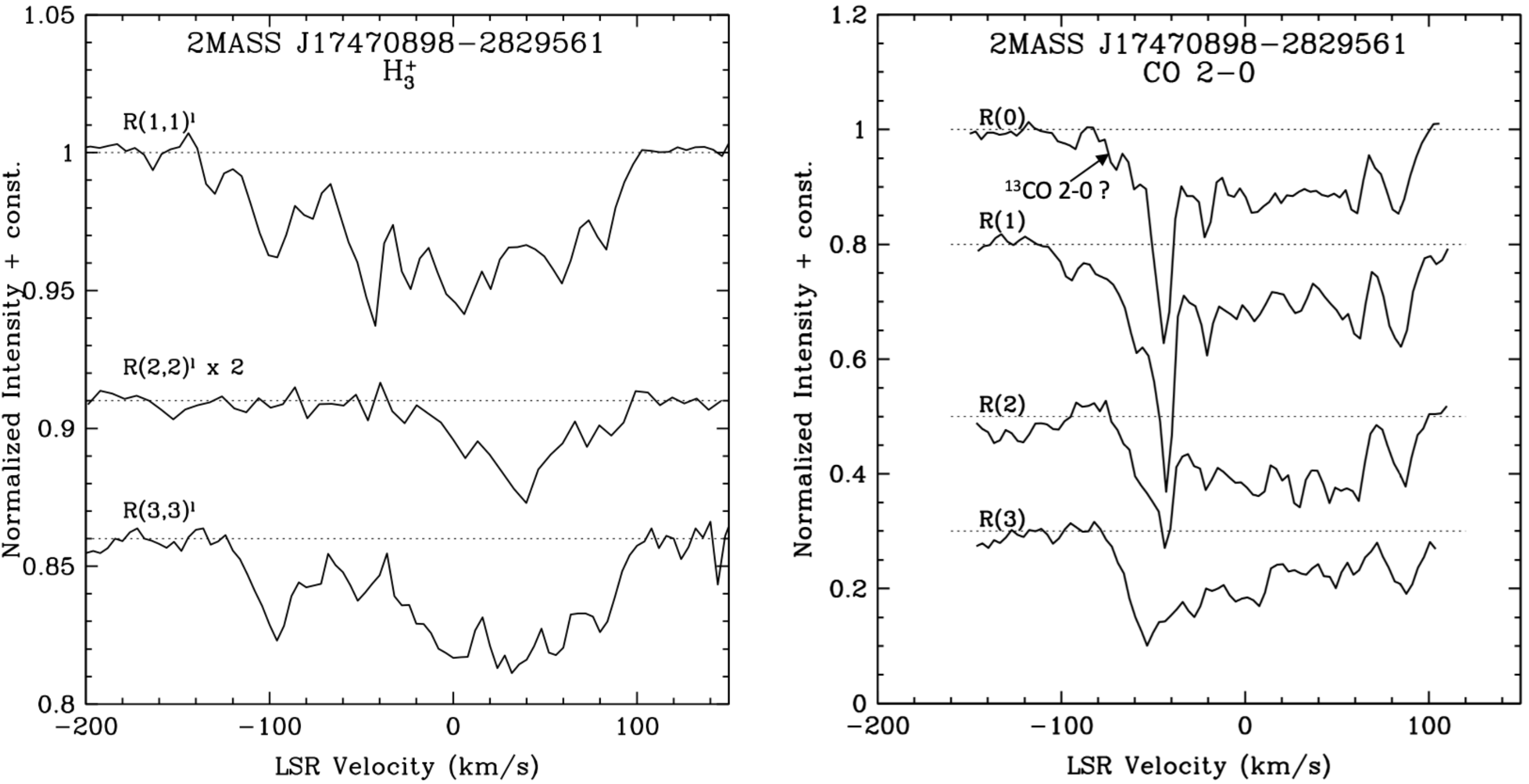}
\caption{Left panel: velocity-resolved spectra of three key lines of H$_3^+$: $R$(1,1)$^{l}$,  $R$(2,2)$^{l}$, and $R$(3,3)$^{l}$ at at 3.7155 $\mu$m, at 3.6205 $\mu$m, and 3.5337 $\mu$m, respectively. Right panel: four lowest $R$-branch lines, $R$(0) - $R$(3), of the overtone vibration- rotation band of $^{12}$C$^{16}$O (right) at 2.3453, 2.3433, 2.3413, and 2.3393 $\mu$m respectively. Wavelengths are in vacuo. The continuum level for each spectrum is denoted by a dotted line. The wavelength of the (marginally detected) $2-0$ band head of $^{13}$C$^{16}$O is indicated by the arrow. Noise can be estimated by the fluctuations at both edges of each H$_3^+$ spectrum and at the left edges of the spectra of the CO lines.}
\label{prevspec}
\end{figure}

\begin{table*}
\centering
\caption{New Spectroscopy of 2MASS $J$17470898$-$2829561\label{t1}}
\begin{tabular}{clcccccccc}
\hline\hline
Date & Program & Telescope & Instr & Exposure & Slit Width & Wavelength & R & Tell. Std. & Airmass \\   
        &                  &                 &          & sec & arcsec & $\mu$m  & $\lambda$/$\Delta\lambda$ &  & 2M1747, tell. std \\
\hline
20100517 & 2010A-8 & IRTF & SpeX & 960 & 0.80 & 2.86-5.30 & 900  & HIP 84848 & 1.51, 1.57 \\
20100722 & 2010A-C-3 & Gemini S & Phoenix & 2160 & 0.34 & 3.613-3.625 & 50,000 & HR 6879 & 1.25, 1.26 \\
20100726 & 2010A-C-3 & Gemini S & Phoenix & 480 & 0.34 & 4.637-4.660 & 50,000 & HR 6165 & 1.04, 1.10 \\
20160909 & 2016B-FT-12 & Gemini N & GNIRS & 800 & 0.45 & 2.19-2.37 & 4,000 & HIP 89622 & 1.66, 1.65 \\
20160909 & 2016B-FT-12 & Gemini N & GNIRS & 480 & 0.20 & 3.03-4.02 & 950 & HIP 89622 & 1.52, 1.59 \\
\hline \hline
\end{tabular}
\end{table*}

\section{Previous Spectroscopy}

\cite{geb10} found that 2M1747 has a steeply rising $K$-band continuum, with multiple weak, CO first overtone absorption bands. These imply intrinsically strong photospheric absorption that is being heavily veiled by continuum emission from warm circumstellar  dust. Their spectrum also contained a prominent dip centered at the wavelength of the CO $2-0$ band center, which encompasses lines arising from rotational levels $J \leq 3$. The intrinsically strong CO absorption bands indicated that 2M1747 is a red giant or supergiant, while the the dip at CO $2-0$ band center indicated that much cooler CO is also present in foreground interstellar gas on the line of sight. The heavy veiling by warm dust makes the source suitable for spectroscopy of H$_3^+$ at longer wavelengths (see Section 5.1) where the dust continuum is much more dominant. In the course of the \cite{geb19} $K$-band survey spectra were obtained of several other bright and highly reddened stars with sightlines close to that of 2M1747 and thus in the general direction of Sgr B. All of them were found to be red giants with little or no veiling of their heavily reddened photospheric spectra. Because of their prominent and extensive photospheric absorption features, they were judged unsuitable for study of the relatively weak lines of H$_3^+$. At present, 2M1747 is the only known star straightforwardly usable for spectroscopy of interstellar H$_3^+$ and CO along a line of sight toward and extending into the Sgr B molecular clouds.  

Figure~\ref{prevspec} shows high-resolution spectra of three vibration-rotation lines of H$_3^+$ and four low-lying CO $R$-branch lines. With the exception of the $R$(2,2)$^{l}$  line of H$_3^+$, which is the average of an unpublished spectrum (see Table 1) and a spectrum in \citet{got11}, these spectra previously appeared in \cite{geb10} and in \cite{oka19}. The H$_3^+$  $R$(1,1)$^{l}$ and  $R$(3,3)$^{l}$  lines and each of the CO lines reveal a multitude of absorption components extending continuously over a velocity range of $\sim$200 km s$^{-1}$. The similar strengths of the $R$(1,1)$^{l}$ and  $R$(3,3)$^{l}$  absorptions over that wide range of velocities imply that almost all of the absorbing gas is at temperatures of $\sim$200 K, while the simultaneous presence from the highest positive velocities to low negative velocities of prominent absorption in the $R$(2,2)$^{l}$ line indicates that much of the gas at those velocities is dense.  As demonstrated later in this paper, the wam dense positive-velocity gas is virtually entirely located within Sgr B1. The significance of these H$_3^+$ lines as diagnostics of gas density and temperature has been presented in several earlier papers \citep[e.g.][]{oka05,oka19,oka20,mil20}.

Two exceptions to the absorbing gas being warm are (1) the prominent sharp absorptions centered at $-$43 km s$^{-1}$, present in the $R$(1,1)$^{l}$ line and in the three lowest-lying CO overtone band lines, and (2) the sharp but weaker absorptions at $\sim$$-$25 km s$^{-1}$ in the same lines. These sharp absorption components are absent from the $R$(3,3)$^{l}$  line, indicating a much lower temperature for much of the gas at those velocities. The velocity of $-$43 km s$^{-1}$ is characteristic of the cold dense gas in the foreground 3 kpc spiral arm, which at that location produces strong absorption in the 6 cm line of  H$_{2}$CO at the same radial velocity \citep{meh95}.  The absorptions at $\sim-$25 km s$^{-1}$ are probably due to gas in the 4.5 kpc arm and can be seen on many other sightlines toward stars in the CMZ \citep{oka19}. CO in dense gas associated with the Local Arm may contribute in part to the absorptions near 0 km s$^{-1}$ in these lines, but does not produce sharp features, as have been found, e.g., by \citet{oka19} on some CMZ sightlines. 

Absorption features from the 3 kpc and 4.5 kpc arms are absent in the CO $2-0$ $R$(3) line (Figure 2, right panel, bottom trace), presumably because in addition to the low kinetic temperatures of the clouds in these arms, the gas densities of the clouds are not high enough to maintain LTE populations in the $J$ = 3 level. For CO the critical density against radiative decay from the $v$ = 0, $J$ = 3 level is $\sim2 \times$ 10$^{4}$ cm$^{-3}$ \citep{yan10}.

At negative velocities the presence of prominent absorption in both the $R$(1,1)$^{l}$ and $R$(3,3)$^{l}$ lines, coupled with the absence of absorption in the $R$(2,2)$^{l}$ line indicates that at those radial velocities the sightline contains warm and diffuse gas. Clearly this gas is part of the diffuse gas expanding radially from the center of the CMZ observed on many other sightlines stretching across the CMZ \citep{oka20}. 

\section{New Observations and Results}

Since obtaining the spectra shown in Figure 2, we have acquired additional infrared spectra of 2M1747 in several key wavelength intervals using infrared spectrographs at three telescopes. The observations are summarized in Table 1. In 2010 we used SpeX at the NASA Infrared Telescope Facility to obtain a 2.9-5.3 $\mu$m spectrum at a resolving power, R, of ~2500. That year we also acquired a high-resolution spectrum of several of the low-lying fundamental band lines of CO and of the $R$(2,2)$^{l}$ of H$_3^+$  at 3.6205  $\mu$m (see Figure 2), using Phoenix at Gemini South. In 2016 we used Gemini North and its facility near-infrared spectrograph, GNIRS, to remeasure the spectrum of a portion of the $K$ band, at much higher resolution than the 2008 spectrum. In addition, we acquired a medium resolution $3-4$ $\mu$m spectrum using GNIRS.  In all cases the spectra were obtained in the standard ABBA manner with nods along the slit of a few arcseconds between the A and B positions.
 
Data reductions, employing IRAF and Figaro packages were also standard. Each reduction included the combining of individual spectral frames, flat-fielding, spectrum extraction, spike removal, cross-correlating and shifting the B spectra to align them in wavelength with the A spectra prior to combining them, wavelength calibration using an arc  lamp or telluric lines, cross-correlating the source spectrum with that of the standard star, shifting the source spectrum to align with that of the standard star, and finally ratioing and conversion of the ratioed spectra to flux densities. For the $L$- and $M$-band spectra, hydrogen Pfund series absorption lines in the spectrum of the standard star were removed by interpolating across them prior to ratioing.  Because of variable sky conditions during the observations and the use of narrow slits, the accuracy of the absolute flux scale, where shown, is $\pm30$\%.

\begin{figure}[]
\includegraphics[angle=-0,width=0.46\textwidth]{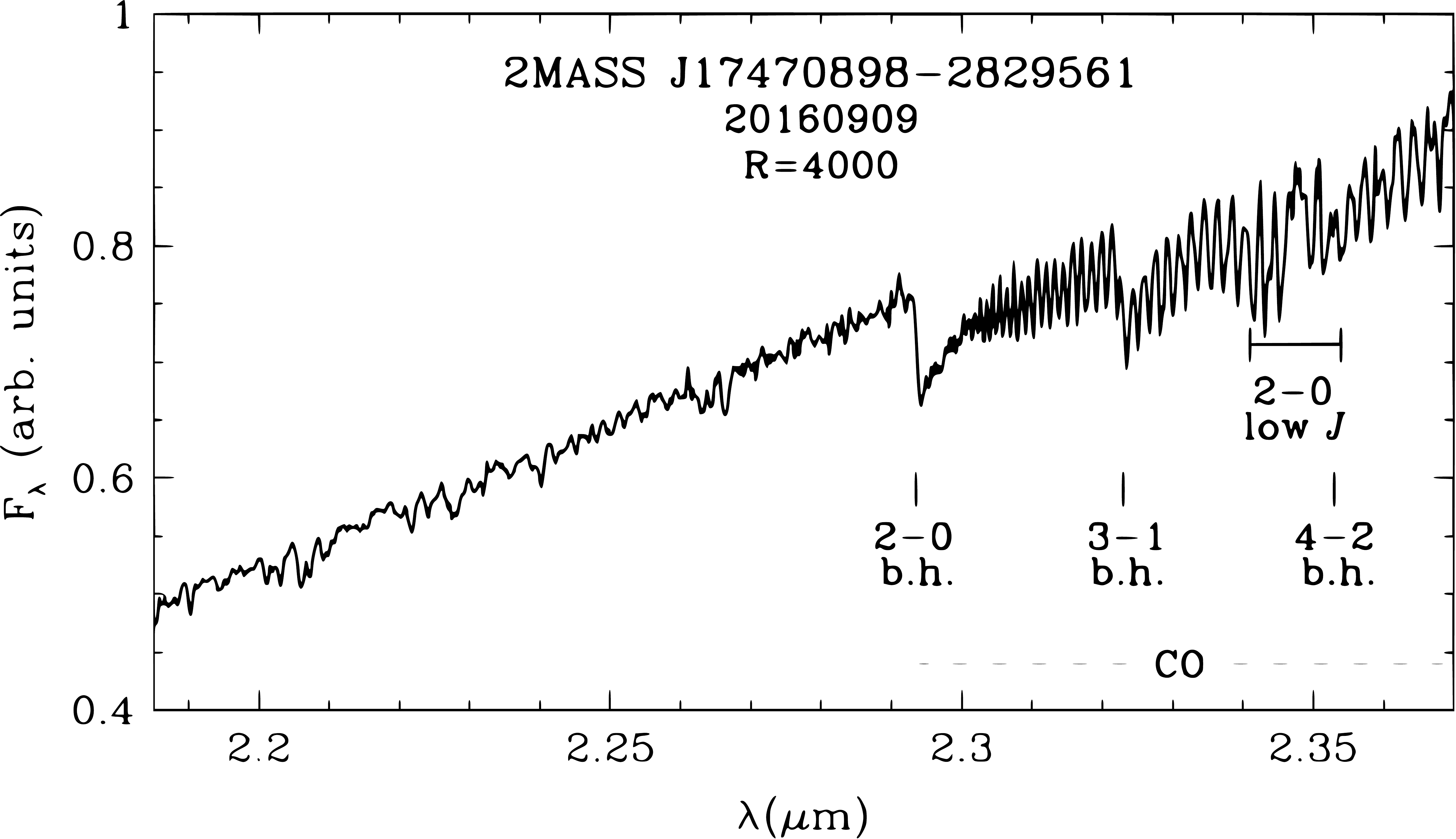}
\caption{Spectra of 2M1747 in a long wavelength portion of the $K$ band obtained at Gemini in 2016. The wavelengths of the first three band heads of $^{12}$C$^{16}$O and the narrow wavelength interval containing lines of the $2-0$ band from low lying rotational levels are indicated.}
\label{kspec}
\end{figure}

\subsection{$K$-band Spectroscopy}

Figure 3 displays the medium-resolution $K$-band spectrum of 2M1747. In comparison to the five times lower resolution spectrum 
in \cite{geb10}, which was obtained in 2008, much more detail is seen and in particular, many of the lines of CO are now resolved from one another. After accounting for the difference in spectral resolution, the depths of the CO bands are approximately 1.5 times greater in the 2016 spectrum, indicating either decreased veiling or a change in the surface gravity of the stellar photosphere.  

The increased strengths of individual absorption lines on either side of the CO $2-0$ band center at 2.3473 $\mu$m relative to neighboring lines is apparent in the Figure 3. It is particularly obvious on the short wavelength ($R$-branch) side of the band, which is as expected due to the greater intrinsic strengths of the low-$J$ lines relative to those on the $P$-branch. In the hot and dense photosphere of the star the CO lines from low-lying rotational levels weaken with decreasing $J$. In contrast, as pointed out in the previous section, only the lowest few rotational levels in the ground vibrational state are populated at low temperature and low density. One may therefore conclude from the spectrum that CO in gas at much lower temperature than gas in the photosphere of the star is the major contributor to the low-$J$ lines. This is also readily apparent from the CO spectra in Figure 2. 

The photospheric $2-0$ band head of $^{13}$C$^{16}$O at 2.3448 $\mu$m is not evident in Figure 3, although there is an indication of its presence in the higher-resolution spectrum in Figure 2 (right panel, top trace). The resolution of the spectrum in Figure 3 is too low to derive an astrophysically significant value of or lower limit to the ratio $^{12}$C/$^{13}$C. See Section 4.2 for somewhat stronger constraints on the value of $^{12}$C/$^{13}$C.

The low-resolution $K$-band spectrum in \cite{geb10} contains a broad absorption feature at $2.26-2.27$ $\mu$m. In the 2016 spectrum (Figure 3) an absorption is also present, but it is much weaker and can be attributed to a triplet of photospheric Ca lines. Other weak absorption features that are present in the 2016 spectrum also are attributable to atomic lines commonly seen in late-type stars \cite[e.g., see][]{wal97}. Examination of the data from 2008 indicates that the $2.26-2.27$ $\mu$m feature there is probably spurious, and is due at least in part to partially corrupted files in the raw data files.

\begin{figure}[]
\includegraphics[angle=-0,width=0.46\textwidth]{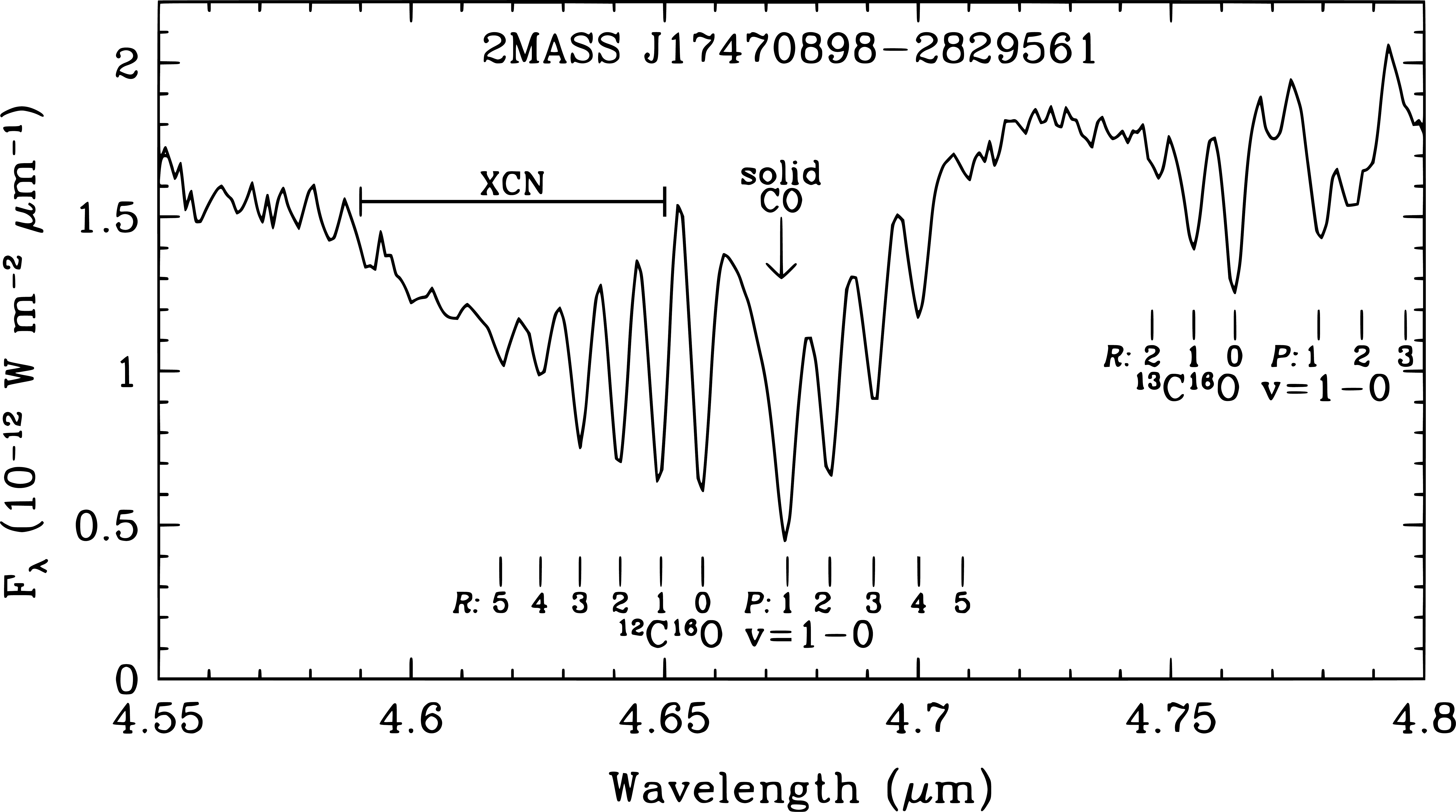}
\caption{Spectrum of 2M1747 in part of the $M$ band, observed by IRTF/SpeX at R=900. Rotation-vibration lines in the 
$P$ and $R$ branches of the fundamental bands of $^{12}$C$^{16}$O and $^{13}$C$^{16}$O are indicated, as are the locations of the solid-state ``XCN" and CO bands. The noise in the spectrum can be judged by the fluctuations at wavelengths less than 4.58 $\mu$m and in the interval 4.71-4.74 $\mu$m.}
\label{mlowspec}
\end{figure}

\subsection{$M$-band Spectroscopy: Low Resolution}

The R=900 spectrum of 2M1747 in the short wavelength half of the $M$ band is shown in Figure 4. The spectrum is dominated by fundamental band lines of CO.  It resembles the medium-resolution spectra taken by the Short Wavelength Spectrometer on the Infrared Space Observatory (ISO SWS) towards stars in the Central Cluster surrounding Sgr A* \citep[][see their Fig. 1]{lut96,mon01} and the spectra of individual stars in that cluster obtained by \cite{mou09}. The major differences are the greater strengths of the CO absorption lines toward 2M1747 and the absence toward 2M1747 of the H I Pf$\beta$ and Hu$\epsilon$ lines, which are emitted by ionized gas in the H II region excited by the hot stars in the  Central Cluster. 

The fundamental band $P$- and $R$-branch lines of $^{12}$C$^{16}$O toward 2M1747 are clearly detected in Figure 4 as individual spectral features out to $P$(4) and $R$(5) and produce significant absorption somewhat beyond those $J$ values, while lines of $^{13}$C$^{16}$O are detected up to $R$(2) and $P$(2).  The absence of absorption from higher rotational levels implies that, as in the case of the low-$J$ lines of the overtone band, the absorption lines arise in interstellar gas rather than in the photosphere of 2M1747. Because of the strong dust continuum of 2M1747, evident from both photometry and spectra, and the presence of such deep fundamental band lines of interstellar CO, photospheric CO absorption in 2M1747, although detected at low resolution in the $K$ band, cannot be seen in the $M$ band in Figure 4. 

Even at R=900 (corresponding to a velocity-resolution of ~330 km s$^{-1}$), the fundamental band absorption lines observed toward 2M1747 are remarkably deep. Their equivalent widths are typically a few times greater than those observed by ISO SWS toward Sgr A* \citep{mon01} or toward individual sources in the central parsec \citep{mou09}. As can be seen in Figure 5 the fundamental band lines from low rotational levels are strongly saturated, and thus the several times greater equivalent widths of the CO fundamental band lines toward 2M1747 correspond to a much larger difference between the CO column densities on the sightlines toward 2M1747 and Sgr A*. This indicates that toward 2M1747 much of the CO absorption toward 2M1747 occurs within the CMZ, a conclusion also reached in Section 2 based only on the high-resolution spectra of low-$J$ CO overtone band lines. The strengths of the detected $^{13}$C$^{16}$O lines are typically one-third those of the corresponding $^{12}$C$^{16}$O lines.  Because the measured value of $^{12}$C$^{13}$C in Sgr B is 24 $\pm$ 7 \citep{hal17}, this also implies on its own that large portions of the profiles of the low-$J$ $^{12}$C$^{16}$O lines are saturated.
 
Also apparent in Figure 4 is the presence of two well-known solid-state absorptions. The solid CO band centered near 4.674 $\mu$m is responsible for the absorption in the $^{12}$C$^{16}$O $P$(1) and $P$(2) lines appearing deeper relative to the low-$J$ $R$-branch lines than they do for $^{13}$C$^{16}$O.  The broad convex shape of the pseudo-continuum centered at 4.62 $\mu$m is due to the ``XCN" band, often attributed to the OCN$^{-}$ anion \citep[][but see also \citet{pon03}]{dem98}, found within many dense clouds in the Galactic plane \citep[see, e.g.][and references therein]{pen99}. For a general review of these and other infrared solid-state absorption features see \cite{boo15}.

\begin{figure}[]
\includegraphics[angle=0,width=0.47\textwidth]{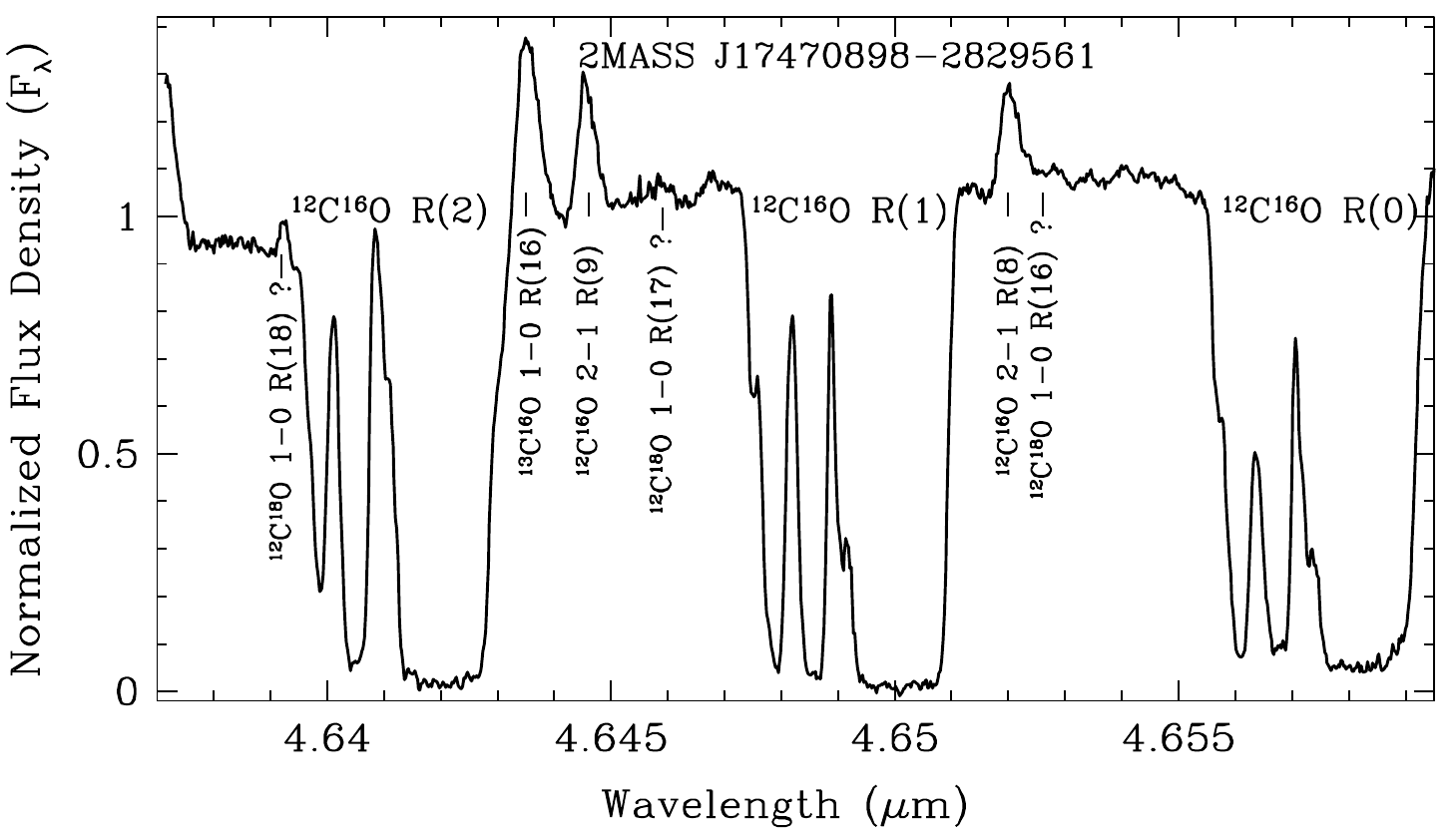}
\caption{High-resolution spectrum covering the lowest three lines of the $R$-branch of the fundamental band of $^{12}$C$^{16}$O, observed by Phoenix at Gemini South.}
\label{mhiresspec}
\end{figure}

\begin{figure}[]
\centering
\includegraphics[angle=-0,width=0.28\textwidth]{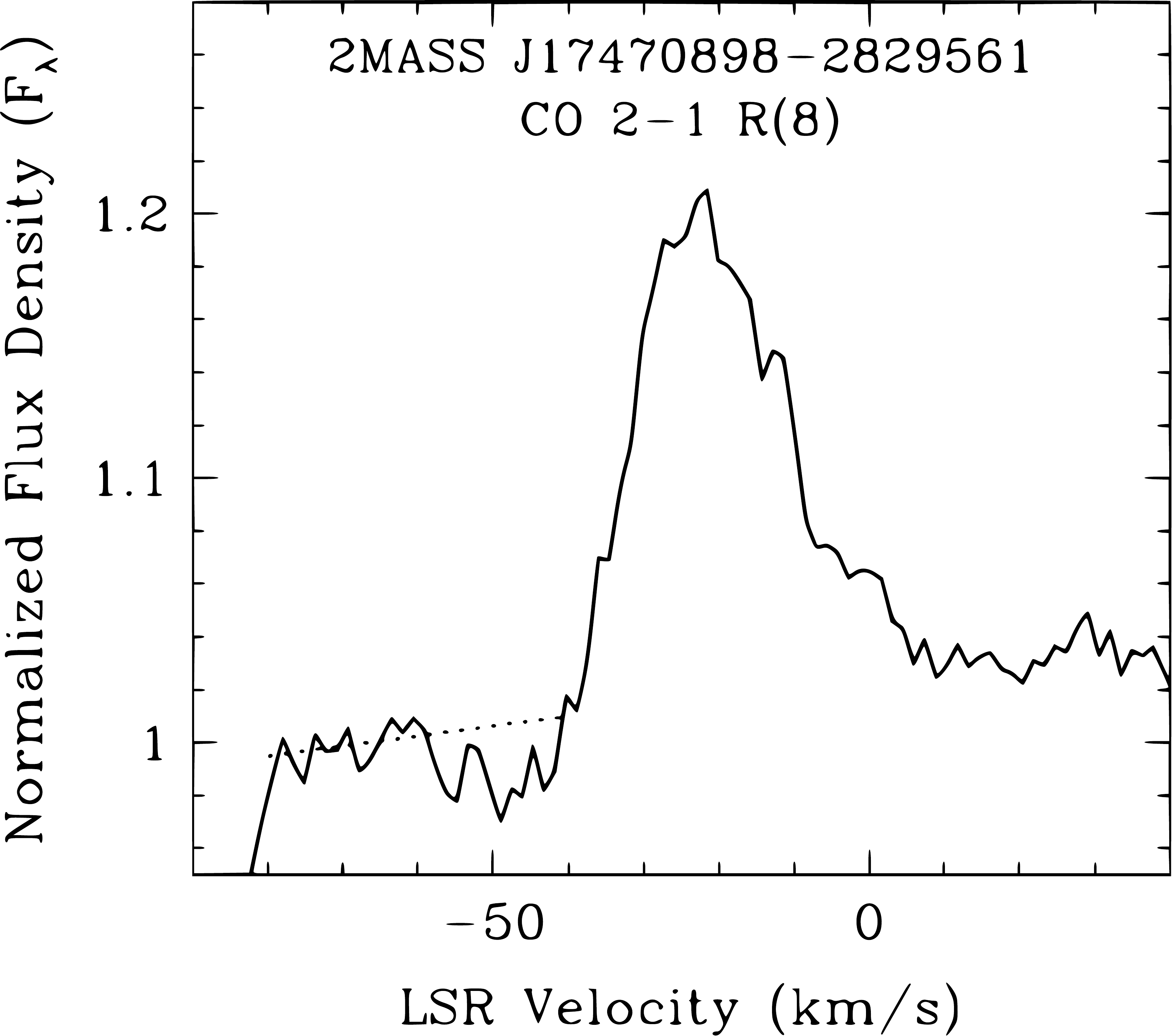}
\caption{Velocity profile of the $2-1 R$(8) line of $^{12}$C$^{16}$O at 4.652 $\mu$m. The dotted line indicates the sloping pseudo-continuum, which is affected by the wing of the strong 1$-$0 $R$(1) absorption line of $^{12}$C$^{16}$O (see Figure 5).}
\label{pcyg}
\end{figure}

\subsection{$M$-band Spectroscopy: High Resolution}

The high-resolution spectrum in Figure 5 provides more detailed information about the CO fundamental band lines. The spectrum covers the wavelengths of the three lowest $R$-branch lines of the $^{12}$C$^{16}$O fundamental, whose absorptions are impressively strong and broad and have similar complex profiles. Five emission lines are also present in the spectrum. Three of them are in the $v = 2-1$ band of $^{12}$C$^{16}$O, one of which, the $R$(10) line, is only partially in the observed spectrum at its short wavelength edge. The  fourth emission feature is a $1-0$ band  transition of $^{13}$C$^{16}$O. The wavelength of the fifth and much weaker emission line, at 4.6392 $\mu$m, matches the wavelength of the $1-0$ $R$(18) line of $^{12}$C$^{18}$O. However, the presences of the $R$(17) (4.6459 $\mu$m) and $R$(16) (4.6526 $\mu$m) lines of this isotopomer, which should be similar in strength to the $R$(18) line, are less clear and thus we regard the detection of $^{12}$C$^{18}$O as marginal. 

Although the strong absorption lines are due to interstellar CO, the emission lines, which originate in excited vibrational states, must be produced in hot gas associated with the star but outside of its photosphere, as discussed below. Lines in the  $v = 1-0$ bands of $^{12}$C$^{16}$O and $^{13}$C$^{16}$O in this narrow interval normally would be present in emission along with the ones that are observed, but their wavelengths coincide with and are masked by the deep and broad interstellar $^{12}$C$^{16}$O absorption lines that dominate the spectrum. The same is true of lines of the $v = 2-1$ band of $^{13}$C$^{16}$O. 

The relative strengths of the emission lines of the  $^{12}$C$^{16}$O and $^{13}$C$^{16}$O isotopomers are similar. Even though the $^{12}$C$^{16}$O emission lines orginate in the $v = 2$ level and the $^{13}$C$^{16}$O line is from $v = 1$, their similar strengths still indicate that the isotopic ratio $^{12}$C/$^{13}$C is low, which is consistent with 2M1747 being evolved. 

Close examination of the $v = 2-1$ lines of $^{12}$C$^{16}$O and the possible $^{12}$C$^{18}$O $1-0$ $R$(18) line reveals that they have P Cygni profiles, with weak absorptions that are blueshifted relative to the much stronger emissions. These profiles indicate the presence of an outflow from the star. The absorption component in the $1-0~R$(16) line of $^{13}$C$^{16}$O near 4.643 $\mu$m, if present, is masked by the strong interstellar absorption of the $^{12}$C$^{16}$O $1-0$ $R$(2) line. However, the full P Cygni profiles of the $2-1$ $R$(8) and $R$(9) lines of $^{12}$C$^{16}$O are unobscured. The former, shown in detail in Figure 6, extends from $v_{LSR}$ $\gtrsim$ 0 km s$^{-1}$ to $-$60 km s$^{-1}$, suggesting that the stellar radial velocity is approximately $-$30 km s$^{-1}$ (LSR). This is consistent with the findings of \cite{lin92}, who determined a stellar radial velocity of $-$32 km s$^{-1}$ (LSR) and an expansion velocity of 32 km s$^{-1}$ for the OH line-emitting shell.  

The negative radial velocity of 2M1747 is somewhat surprising in view of the positive radial velocities of the Sgr B1 cloud complex. One can only speculate as to the explanation, because little or nothing is known about the motions of other stars within the complex. One possibility is that this highly unusual star was ejected in a direction toward the front of Sgr B1 from a more embedded cluster of massive stars and thereby acquired a non-representative radial velocity.

\subsection{\it Excitation of the Infrared Emission Lines}

We now discuss the excitation of the CO $v$ = 1 and 2 rovibrational levels that are observed to produce the emission lines in Figures  5 and 6.  The  $v$ = $1-0$ emission line of $^{13}$C$^{16}$O is from  $J$ = 16 and judging from its strength $v$ = $1-0$ line emission extends at least to somewhat higher $J$.  The $v$ = 0, $J$ = 16 level is 750 K above ground. Significant thermally excited populations of $v$ = 0, $J$ $\sim$ 16 levels require densities $n$ $\gtrsim$ 10$^6$ cm$^{-3}$ \citep{yan10} and temperatures of at least several hundred kelvin. Both appear to be well within the range of values for much of the gas in the circumstellar envelope of an OH/IR star \citep[see Figs. 2 and 4 of][]{gol76}. Although this suggests that resonant scattering of infrared photons off of CO molecules in the ground vibrational state could contribute to the $1-0$ line emission, a much stronger blueshifted absorption feature would likely be present  if that were the dominant cause of the emission. We conclude that this mechanism probably is only a minor contributor to the observed P Cygni line profiles.

Collisional excitation of $v$ = 0 levels to the observed $v$ = 1 levels, which are more than 3,000 K above the ground vibrational state, requires considerably higher temperatures than several hundred kelvin. Such temperatures, perhaps up to $\sim$2,000 K, are also expected to be present in circumstellar gas closest to the stellar photosphere \citep{gol76}. We tentatively conclude that this is the cause of most of the emission seen in the $v = 1-0$ lines.

The $^{12}$C$^{16}$O $v$ = 2 rovibrational levels that produce the $v$ = $2-1$ lines in Figures 5 and 6 could be excited radiatively from $v$ = 0 and 1 levels by resonant absorption of 2.3 $\mu$m and 4.6 $\mu$m photons, respectively.  If 4.6 $\mu$m absorption dominated, one would again expect to see P Cygni profiles with deep absorptions in the $2-1$ lines, which are not observed. However, absorption by 2.3 $\mu$m photons, produced copiously in the stellar photosphere and the hot innermost part of the circumstellar envelope, could be masked by the complex CO overtone band absorption profiles in Figure 2. In addition, those absorptions do not result in resonant emission in  $\Delta$$v$ = 2 lines since the radiative decay of $v = 2$ levels proceeds almost entirely via the much faster $\Delta$$v$ = 1 transitions. Thus radiative excitation of the $v = 2$ levels by 2.3 $\mu$m photons appears to be a viable mechanism for producing $v = 2-1$ emission. 

One also must consider populating CO $v$ = 2 levels from $v$ = 1 by collisions. That requires a jump in energy of $\sim$3,000 K, similar to the energy gap between $v$ = 0 and $v$ = 1, for any value of $J$.  For this process to contribute significantly to the $2-1$ line emission, densities $\gtrsim$ 10$^{10}$ cm$^{-3}$ are needed to maintain CO in the $v$ = 1 states against radiative relaxation to $v$ = 0. For OH/IR stars with high mass loss rates, those required values of temperature and density are expected to exist in circumstellar gas close to the photosphere \citep{gol76}. We conclude that resonant scattering of 2.3 $\mu$m photons and/or  collisional excitation, both operating in the hot and dense inner part of the envelope, could lead to the observed emission in the $v$ = $2-1$ lines. Observations and analysis of a wider range of rovibrational lines in 2M1747 could constrain the relative contributions of these processes.

Finally, we note that excited vibrational levels of CO also can be radiatively populated as a result of absorption by CO of ultraviolet (UV) radiation followed by vibrational cascades, as proposed by \citet{bro13} for protoplanetary disks. Because 2M1747 is a cool giant, however, UV excitation is an unlikely mechanism here.

\subsection{Detailed examination of the profiles of H$_3^+$ and CO infrared and millimeter lines}

Figure 7 shows the velocity profile of the CO fundamental band $R$(1) absorption line (the central strong absorptions in Figure 5). Also in the figure and plotted on the same normalized scale for comparison are the $R$(1) line of the 130 times less optically thick first overtone band and the $R$(1,1)$^{l}$ line of  H$_3^+$ (magnified by a factor of 4). In addition, the figure includes at top the velocity profile of the 2.6 mm $J$ = 1$-$0 line of $^{12}$C$^{16}$O at a position 14\arcsec distant from 2M1747, observed in the \cite{oka98} GC survey at the 45 m telescope of Nobeyama Radio Observatory, which used a  17\arcsec diameter  beam (0.66 pc at the distance of the GC) and a 34\arcsec grid spacing. 

The spectrum of the pure rotational CO line is dominated by emission at positive velocities, which are as high as + 110 km s$^{-1}$. The wide range of velocities is characteristic of the Sgr B1 and Sgr B2 cloud complexes \citep{sco75}. The velocity profile of the line is surely affected by radiative transfer effects; thus a detailed comparison of it with the profiles of the infrared absorption lines is probably not meaningful, apart from noting that most of the absorption equivalent width in each of the latter is in the same range of positive velocities as the emission in the pure rotational line. 

\begin{figure}[]
\includegraphics[angle=0,width=0.46\textwidth]{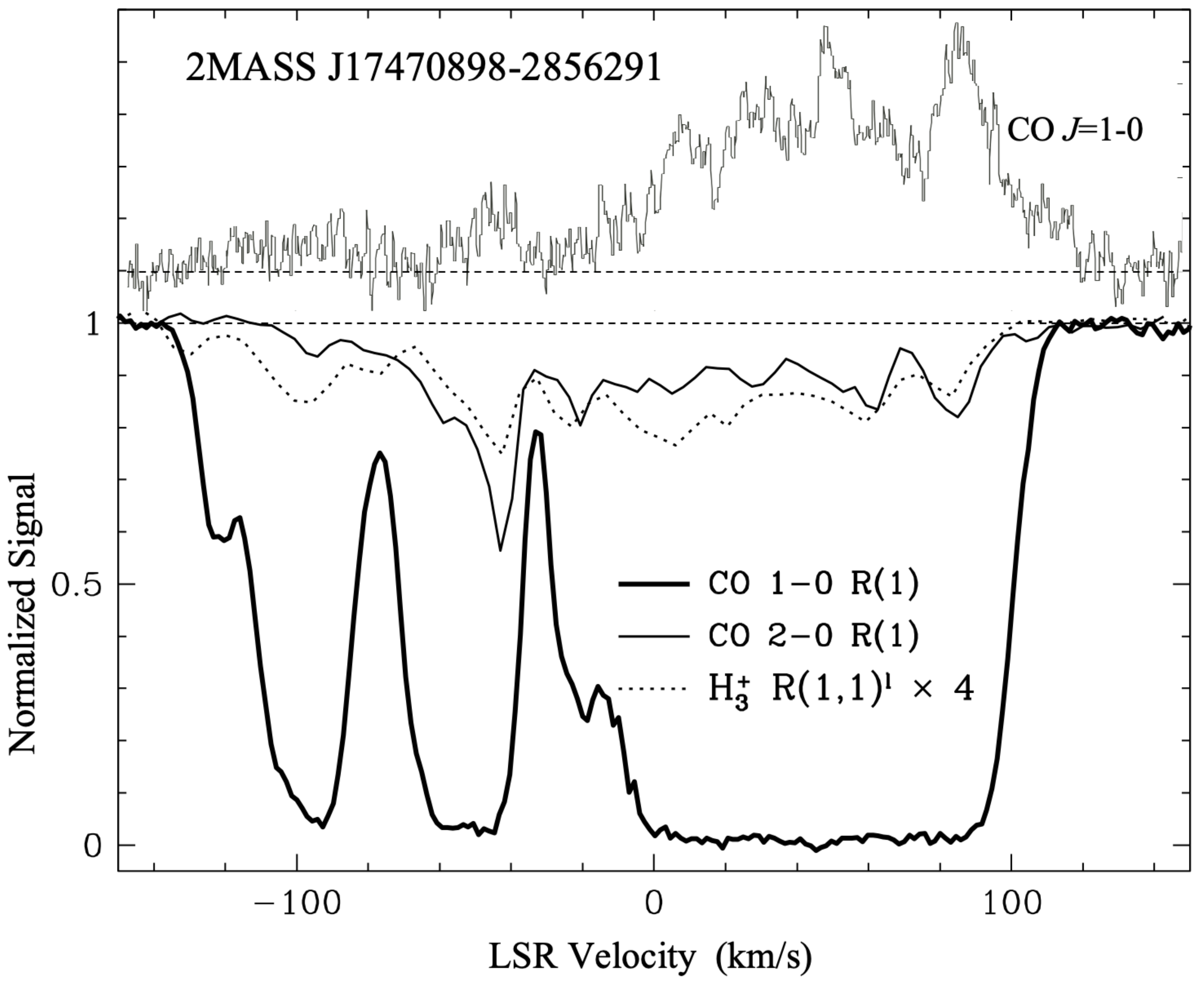}
\caption{Top trace: spectrum of the CO $J$ = 1$-$0 pure rotational line at the closest position to 2M1747 obtained by \cite{oka98} and kindly made available by Tomoharu Oka. The zero level is indicated by the upper dashed line. Lower three traces: normalized spectra  of the $R$(1) lines of the fundamental and first overtone band of CO  (thick and thin lines, respectively) and the $R$(1,1)$^{l}$ line of H$_3^+$ (dotted line, magnified by a factor of 4). }
\label{m_vel}
\end{figure}

We now look in detail at the infrared line profiles in Figure 7. One would expect that at wavelengths where the depth of the absorption in the CO $2-0$ $R$(1) line is more than several percent the absorption in the $1-0$ $R$(1) line would be nearly total. This is indeed the case apart from a narrow velocity interval centered at $-20$ km s$^{-1}$. The two infrared CO lines illustrate the difficulty in determining cloud type and temperature in the GC from studies based largely on spectroscopy of the strong fundamental band lines \cite[e.g.][]{geb89,mou19}. This is especially clear at high negative velocities where there is strong absorption in the $1-0$ $R$(1) line, but where the $2-0$ $R$(1) line is very weak and the relative strengths of the H$_3^+$ lines in Figure 2 show that the absorbing gas at those  velocities is warm and diffuse.  

The absorption in the CO $1-0$ $R$(1) line extends from $-130$ km s$^{-1}$  to $+110$ km s$^{-1}$. Its profile contains three ranges of radial velocities within which the optical depth is 3 or higher: 0 km s$^{-1}$ $<  v < +90$  km s$^{-1}$, $-60$ km s$^{-1}$ $<  v < -40$ km s$^{-1}$, and $-105$ km s$^{-1}$ $<  v < -90$ km s$^{-1}$. The first of these (at positive velocities) is the same range as appears at this position in the millimeter-wave CO spectrum in Figure 7. The strong absorptions in the H$_3^+$ $R$(2,2)$^{l}$  and $R$(3,3)$^{l}$ lines in this velocity range (Figure 2), show that this gas is warm and mainly dense. 

The deep CO $1-0$ $R$(1) absorption in the middle velocity range ($-60$ km s$^{-1}$ $<  v <$ $-40$ km s$^{-1}$ ) is due in part to absorption outside of the CMZ by cold dense gas in the foreground 3 kpc arm \citep{oka19}, which is the most prominent absorption component in the both the overtone line of CO and the H$_3^+$ line. Note, however, that warm diffuse gas must also be present in this middle velocity range, because absorption in the $R$(3,3)$^{l}$  line of H$_3^+$ is present while absorption in the $R$(2,2)$^{l}$  line is absent (see Figure 2). Thus absorption by the CO fundamental band line in this velocity range is a superposition of warm diffuse blueshifted gas, within the CMZ (see Section 3) and cold dense gas in the 3 kpc arm. 

Somewhat similarly, the absorption in the CO  $1-0$ $R$(1) line at velocities between $-30$ and 0 km s$^{-1}$ contains contributions from cold dense gas in the 4.5 kpc arm and possibly also in the Local arm as was discussed in Section 3, as well as dense and warm gas associated with Sgr B1. CO in the warm and diffuse gas of the CMZ may also contribute absorption in the fundamental band line at these velocities.  Warm diffuse gas in this velocity range in known to exist on many sightlines across the CMZ \citep{oka20}. On this sightline, however, it cannot be distinguished from the other contributors.

Finally, we conclude that the CO absorption observed at velocities more negative than $\sim-60$ km s$^{-1}$ also arises mainly in warm diffuse gas, not only because in that velocity range absorption in the H$_3^+$ $R$(3,3)$^{l}$ line is present but absorption in the H$_3^+$ $R$(2,2)$^{l}$ line is absent, as was mentioned in Section 3, but also because absorption in the CO overtone lines is relatively weak. Thus, blueshifted warm and diffuse gas extends from near 0 km s$^{-1}$ to $-130$ km s$^{-1}$. It is part of the warm diffuse gas observed by \cite{oka19} and \cite{oka20} toward numerous stars in the CMZ, attributed by them to filling the majority of the volume of the CMZ and undergoing radial expansion from a location near the center of the CMZ.  It includes the gas at the highest velocities, part of the ``Expanding Molecular Ring" (EMR) at the outer edge of the CMZ, first observed at radio wavelengths by \cite{kai72} and \cite{sco72} and analyzed in detail by \cite{sof95} and \cite{sof17}. 

\subsection{$L$-band Spectroscopy}

Figure 8 shows the $3-4$ $\mu$m spectrum of 2M1747, observed at R = 900. The spectrum is an adjoining of the $2.86- 3.04$ $\mu$m portion of the SpeX spectrum and the full GNIRS spectrum (see Table 1). The figure also contains an approximate optical depth spectrum (lower trace), which was generated using a continuum based on a spline interpolation of the spectrum as described below. 

To create the spline we assumed that the observed data points near 3.60, 3.70, 3.78, and 4.01 $\mu$m represent the continuum. This should be a good approximation because no known spectral features occur at those wavelengths and because the slope of the continuum through these data points changes only slightly between 3.60 $\mu$m  and 4.01 $\mu$m.  At wavelengths less than 3.60 $\mu$m several overlapping solid-state bands cover the entire observed range. Based on the analysis of spectra toward  seven stars in the Galactic center \cite{chi02} derived a mean ice-band optical depth spectrum (see their Fig. 5). We have used that spectrum to estimate the continuum level in the 2M1747 spectrum at 3.34 $\mu$m, which corresponds to a local maximum in the observed flux density and which is a wavelength where the wing of the ice band is the only solid-state feature contributing significant absorption. At that wavelength the optical depth of the mean spectrum is 0.4 times the optical depth at the bottom of the ice band.  Finally, to estimate the  continuum level at the short wavelength edge of the spectrum at 2.86 $\mu$m, which is on the short wavelength shoulder of the ice feature, we have made use of the  \cite{lut96} $2.4 - 45~\mu$m spectrum toward Sgr A*, which includes  continuum at wavelengths less than the short wavelength edge of the above absorption feature. Based on it we estimate that the continuum level at 2.86 $\mu$m is 12\%\  higher than the observed flux density. 

\begin{figure}[]
\includegraphics[angle=0,width=0.46\textwidth]{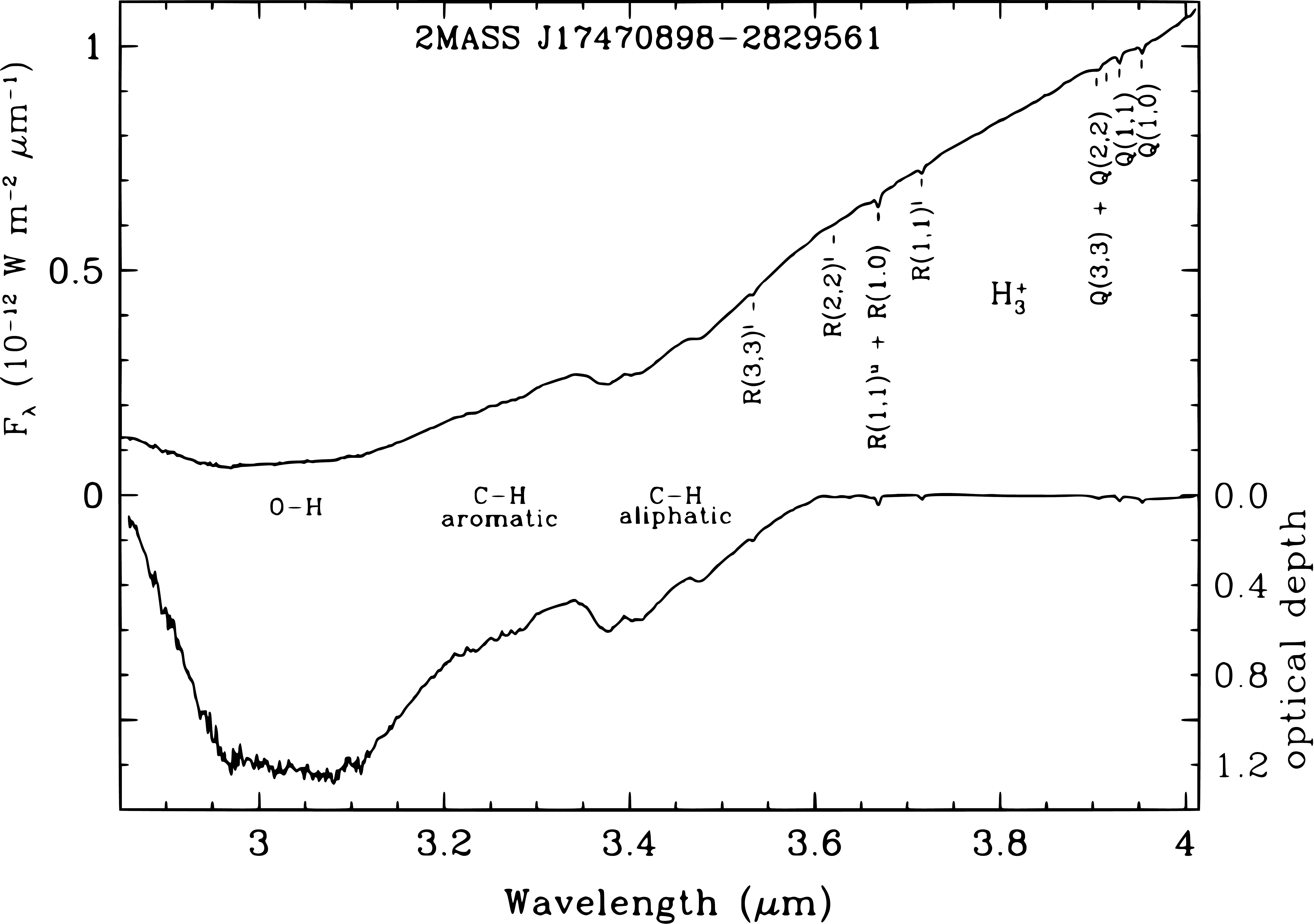}
\caption{Upper trace: $3-4 \mu$m spectrum of 2M1747. Identifications and wavelengths of lines of H$_3^+$  are shown. Lower trace: optical depth spectrum, with scale shown at right edge. The locations of the O-H stretching band and C-H stretching bands are indicated.}
\label{lspec}
\end{figure}

The spectrum in Figure 8 contains six weak but clearly detected narrow absorptions due to individual or blended lines of H$_3^+$. A seventh feature, due to the $R$(2,2)$^l$ line at 3.6205 $\mu$m, is only marginally detected, but is easily seen in Figure 2. Over this rather wide wavelength range and at this rather low spectral resolution the only interstellar gas-phase lines detected toward this highly obscured source are those of H$_3^+$. 

\begin{figure}[]
\includegraphics[angle=0,width=0.46\textwidth]{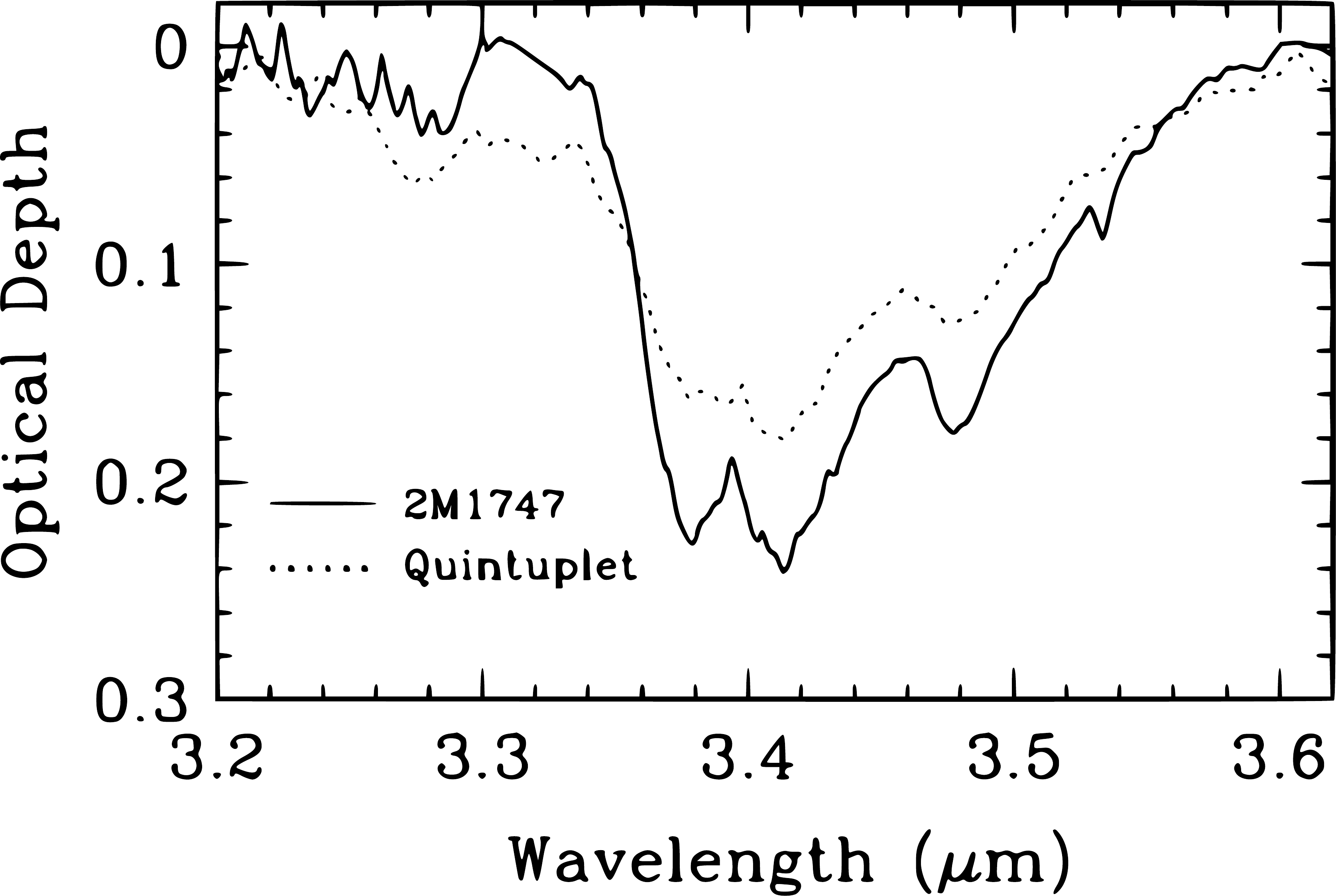}
\caption{Comparison of the optical depth spectra of the 3.4 $\mu$m absorption features toward 2M1747 (this paper) and the Quintuplet Cluster (a weighted average of the spectra in \cite{chi13}, binned to the resolution of the Gemini spectrum). The optical depth of the wing of the H$_2$O ice band, which is superimposed on the 3.4 $\mu$m feature, has been subtracted. The broad and shallow 3.25 $\mu$m feature and three characteristic absorption peaks of the 3.4 $\mu$m feature at 3.38, 3.42, and 3.48 $\mu$m are clearly present. The narrow absorption at 3.53 $\mu$m in the spectrum of 2M1747 is a line of H$_3^+$ (see Figure 8).}
\label{l34spec}
\end{figure}

The spectrum also contains several broad absorption bands, most or all of which have been detected toward other objects in and outside of the GC. Their properties are summarized in Table 2. The strongest absorption feature, covering the $2.86-3.60~\mu$m range with a peak near 3.0 $\mu$m, has been attributed largely to the OH stretching vibration in water ice \citep[e.g.][]{san91,chi00}. In some sources in which the water ice is believed to be fairly pure \citep[e.g.,][] {hod88,geb88,smi89}, as well as in models \citep{chi02} the absorption maximum occurs at $3.05-3.08~\mu$m, with a fairly sharp decrease in the absorption on either side. Toward 2M1747, however, as well as toward other sources in the Galactic center, a broad absorption maximum extends shortward to 2.97 $\mu$m. The additional contribution has been attributed to impurities in the ice, in particular NH$_3$ \citep[e.g.][]{chi00}.

The structured absorption extending from 3.35 $\mu$m to 3.60 $\mu$m, superimposed on the long wavelength wing of the ice band and first observed toward the Galactic center by \cite{soi76}, \cite{jon83}, and \cite{but86}, is commonly referred to as the 3.4 $\mu$m feature. It is known to be ubiquitous in the diffuse ISM, but is not detected in dense molecular clouds. The absorption is generally acknowledged to be due to asymmetric C$-$H stretching modes of CH$_3$ and CH$_2$ groups in saturated aliphatic hydrocarbons \citep{san91}. The three absorption maxima, at 3.38, 3.42, and 3.48 $\mu$m in the spectrum of 2M1747, better seen in Figure 9, not only are characteristic of its profile on lines of sight to other infrared sources in the Galactic center, but are also present in spectra toward diffuse clouds in the Galactic plane \citep{san91,pen94}. The 3.4 $\mu$m feature also has been observed in the interstellar dust of external galaxies and studied extensively in one of them \citep[][and references therein]{dar07}. 

A third, broad and much shallower absorption in Figure 8 extends from 3.22 $\mu$m approximately to 3.30 $\mu$m, and can be seen in more detail in Figure 9. The absorption profile toward 2M1747 resembles that of the mean Quintuplet spectrum of \citet[][the lower trace of their Fig. 4 is reproduced in Figure 9]{chi13}. In particular the maximum absorption in the above spectral interval, at 3.28 $\mu$m, matches a distinct local absorption maximum in their spectra, which they identify as due to the C$-$H stretch in the olefinic constituents of hydrocarbon grains.  However, this vibrational mode is expected to only produce a narrow absorption and cannot account for the entire width of the observed feature. 

A weak absorption first reported by \cite{sel95} toward the young stellar object (YSO) Mon R2 IRS3, and subsequently observed toward several additional YSOs  \citep{bro96,bro99,bre01,har14} also covers the $3.2-3.3$ $\mu$m spectral interval. Both \cite{bro99} and \cite{bre01} suggest that it is due to the C$-$H stretching band in mixtures of isolated aromatic hydrocarbon molecules in cold dense clouds. A second possible identification, ammonium (NH$_4^+$) in icy grain mantles, has also been suggested \citep{sch03}. The optical depth of the feature observed in these sources is typically a few times greater than toward 2M1747 and the Quintuplet stars. Although the signal-to-noise ratio of individual spectra reported by the above observers is low, the absorption appears to be smooth, without a narrow peak at 3.28 $\mu$m. Thus one must entertain the possibility that the observed $3.22-3.30$ $\mu$m absorption observed toward objects in the GC is a superposition of two features, possibly arising in different portions of the line of sight. 

\begin{figure}[]
\includegraphics[angle=0,width=0.46\textwidth]{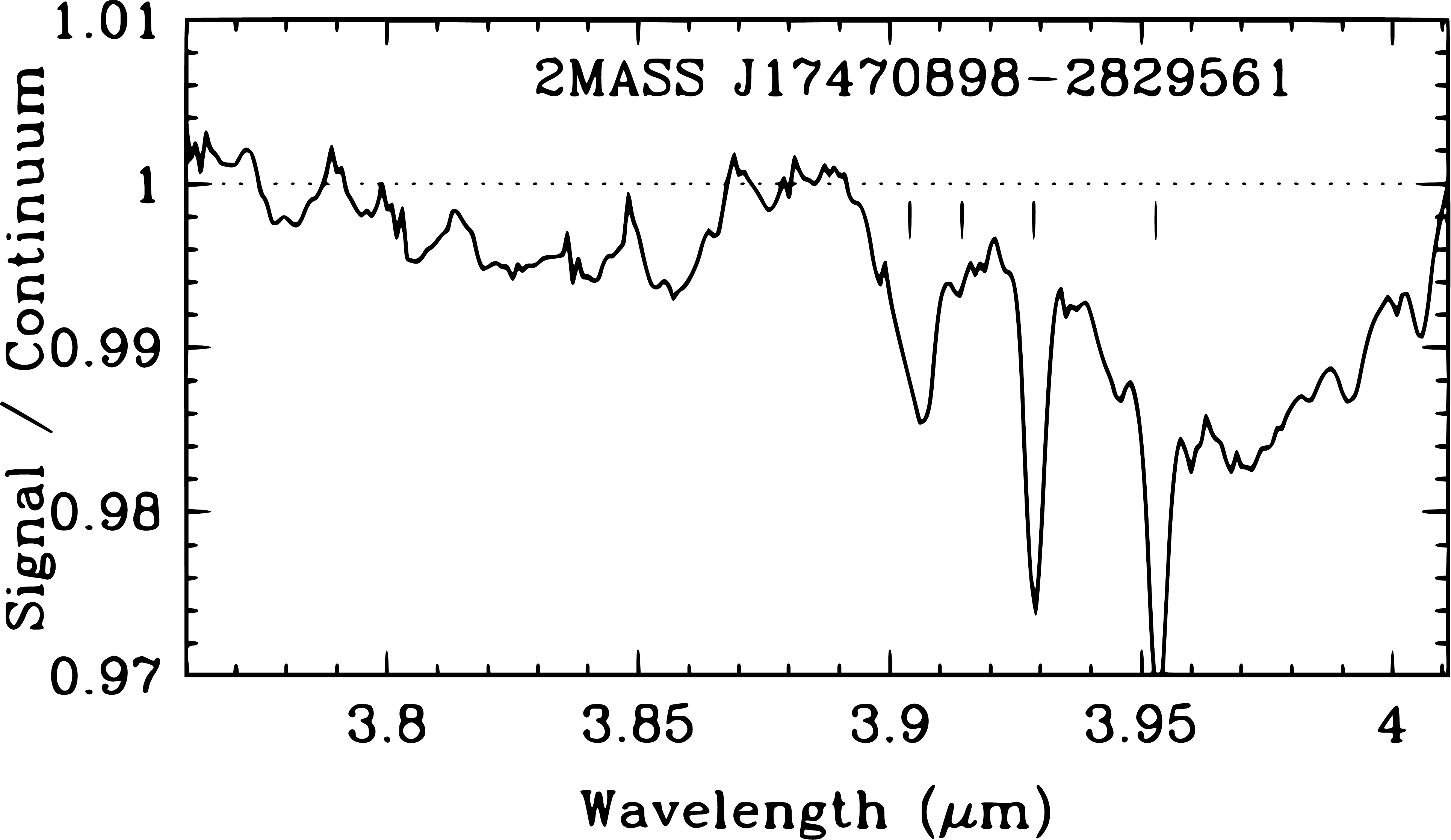}
\caption{Expanded view of normalized spectrum of 2M1747 in Figure 8 showing evidence for broad absorption features in the $3.8-4.0 \mu$m interval. The dotted horizontal line is the assumed continuum level from Figure 8. Short vertical lines denote absorption lines of H$_3^+$ identified in Figure 8.}
\label{lpspec}
\end{figure}

\begin{table*}
\centering
\caption{Absorption Features of Dust Grains Toward 2M1747 \label{t2}}
\begin{tabular}{ccccc}
\hline\hline
Wavelength & Identity &  optical depth &Equiv. Width &  Locations \\   
 microns      &                  & at peak & microns   &        \\
\hline 
$2.8-3.6$ & H$_2$O $+$ impurities  & 1.25 & 0.20 & Sgr B1 dense gas, spiral arm dense gas \\
$3.20-3.30$ & aromatic C$-$H & 0.03 & 0.0018 &  \\
$3.35-3.60$ & aliphatic $-$CH$_{2}$, $-$CH$_{3}$ & 0.24 & 0.029 & spiral arm diffuse gas, CMZ diffuse gas (?)  \\
$3.48-3.60$ & CH$_{3}$OH & $<$0.002 & $<$0.0003 & (not detected) \\
$3.80-3.86$ & $?$ & 0.005 & 0.00027 $\pm$ 0.00005 & \\
$3.92-4.00$ & CH$_{3}$SH $?$ & 0.015 & 0.0010 $\pm$ 0.0002 & \\
$4.585-4.65$ & XCN (OCN$^-$$?$) & 0.2 &0.012 $\pm$ 0.003  & Sgr B1 dense gas (?), spiral arm dense gas (?), CNR (?)\\
$4.66-4.70$ & CO & 0.3 & 0.007 $\pm$ 0.003 & spiral arm dense gas, Sgr B1 dense gas (?) \\
\end{tabular}
\end{table*}

Two absorption bands are present in the $3.8-4.0~\mu$m interval and are more easily seen in Figure 10 than Figure 8. A weak absorption appears to be present from approximately 3.80 $\mu$m to 3.86 $\mu$m; we hereafter refer to it as the 3.83 $\mu$m feature. A stronger absorption feature is clearly present between 3.92 $\mu$m and 4.00 $\mu$m; we hereafter refer to it as the 3.96 $\mu$m feature. Overall the features in Figure 10 crudely resemble those observed previously toward the only infrared sources toward which solid-state features have been detected in the wavelength interval, the YSOs W 33A and AFGL 7009S, which are embedded in dense and cold molecular clouds \citep{geb85,geb91,dar99}. Although detection of the shorter-wavelength absorption feature in W33A was not reported by \cite{geb85} there is an indication of its presence upon close examination of their spectrum. It is more clearly seen in the \cite{dar99} spectra of the above objects.

\cite{geb85} tentatively identified the longer-wavelength feature in the two YSOs as due to solid H$_2$S. Later, however, based on laboratory experiments \cite{dar99} identified the absorptions as due to two overlapping bands of methanol (CH$_3$OH) ice centered near 3.85 $\mu$m and 3.95 $\mu$m. A stronger band of CH$_3$OH ice centered at 3.54 $\mu$m with a full width at zero intensity of ~0.10 $\mu$m is also present toward both YSOs, strengthening their case for identifying the longer-wavelength features as due to frozen CH$_3$OH. The ratios of the optical depths of the features in the two YSOs are $\tau$(3.54)/$\tau$(3.93) $\sim$ 10 and $\tau$(3.54)/$\tau$(3.83) $\sim$ 15.

Toward 2M1747, however, the 3.54 $\mu$m band is not detected (see Figure 9); we estimate $\tau$(3.54)/$\tau$(3.96) $<$ 0.3 and $\tau$(3.54)/$\tau$(3.83) $<$ 1. These limits are more than an order of magnitude less than toward W33A and AFGL 7009S.  The large difference indicates that the neither of the longer-wavelength absorptions toward 2M1747 is due to CH$_3$OH ice. We have considered the possibility that the longer-wavelength feature is due to H$_2$S frozen on dust grains, as had been suggested earlier. The S$-$H stretching mode of solid H$_2$S  has long been known to produce a spectral feature near 3.92 $\mu$m. Laboratory spectra of solid amorphous H$_2$S by \cite{hud18} reveal that the full width at half maximum (FWHM) of this feature is $\sim$ 0.06 $\mu$m. Although the width of the laboratory feature is similar to that observed toward 2M1747, the central wavelengths do not match very well. In dirty ice mixtures the wavelengths of vibrational stretching modes  are often shifted. For example, the well-known band of solid CO at 4.67 $\mu$m is shifted to longer wavelengths by  $\sim$ 0.01 $\mu$m when in a mantle dominated by H$_2$O ice. A considerably larger shift than this would be  needed for the identification of the feature toward 2M1747 as H$_2$S to be viable. Inspection of the laboratory studies of frozen H$_2$S  \citep{moo07,sou19} indicates that such a shift is unlikely.

Perhaps a more plausible identification for the carrier of the 3.96 $\mu$m feature  is methanethiol (CH$_3$SH) ice. The absorption peak of the S$-$H stretch in amorphous CH$_3$SH ice is 3.945 $\mu$m \citep{hud18}, a much closer match to the observed band. It is necessary to determine the size and direction of  wavelength shifts of CH$_3$SH ice in realistic interstellar ice mantles in order to judge whether this is an astrochemically viable identification. At present we have no proposed identification for the much weaker 3.83 $\mu$m feature.

\section{Analysis and Discussion}

\subsection{An Estimated Extinction to 2M1747}

There are two lines of evidence indicating that the bulk of the extinction to 2M1747 arises in dense gas within the CMZ. One is based simply on the H$_3^+$  and CO line profiles in Figure 2, which show that at velocities more positive than $\sim -20$ km s$^{-1}$, where most of the CO absorption occurs  the gas in which the H$_3^+$ and CO are present is warm, indicating that it is located in the CMZ, and that it has a mean density well in excess of the critical density of $\sim$200 cm$^{-3}$ for the (2,2)$^l$ level of H$_3^+$, suggesting that it is mostly found in dense clouds. The second is the huge column density of absorbing CO, $2.7 \times 10^{19}$ cm$^{-2}$ found by \cite{oka19}, derived from the optically thin CO overtone band lines in Figure 2, and what that would imply for the extinction to 2M1747 if even a modest fraction of it were associated with diffuse gas (see Section 2). 

Based on our conclusions that (1) the column density of CO derived from the overtone band is virtually entirely in dense gas and (2) the sharp CO absorptions at $-43$ and $-25$ km s$^{-1}$ are also due  to dense gas (in the 3 kpc and 4.5 kpc arms), we estimate that $\sim$90\% of the column density of CO observed toward 2M1747 ($2.4 \times 10^{19 }$ cm$^{-2}$) is located in the CMZ. Using the ratio of gaseous CO to visual extinction derived from measured strengths of the interstellar H$_2$ quadrupole $S$(0) absorption line and absorption lines of the CO first overtone lines toward stars behind the Taurus Molecular Cloud, \cite{lac17} derived $N_{gas}$(CO)/ $A_V$ $\sim$ $1.5 \times 10^{17}$ cm$^{-2}$. Lacy et al. do not give an uncertainty in this value; we estimate it to be  $0.6 \times 10^{17}$ cm$^{-2}$ (1$\sigma$). We note that the Lacy et al. value is somewhat higher than other reported and derived values \citep{pin10,lac17}; however it is the only direct measurement of that ratio.  Using it the observed column density of gaseous CO toward 2M1747 implies a visual extinction of $180 \pm 72$ mag due to dust in dense clouds. Were a significant fraction of the CO in the line of sight to 2M1747 located in diffuse gas, where the gaseous CO is heavily depleted due to photodissociation, the above formidable value would be larger by $1-2$ orders of magnitude, and would lead to 2M1747 being unphysically luminous, as discussed in Section 5.2. Note that due to the higher metallicity toward the Galactic center \citep[for a recent summary of previous measurements and an estimate see Section A.2 of][]{oka19}, the value of $N$(H)/ $A_V$ there is probably considerably lower than the more local value. However, to first order $N$(CO)/$A_V$ should not vary too much, as the abundances of both dust and CO are enhanced by the higher metallicity.

The contribution of extinction toward 2M1747 due to dust in diffuse gas on its sightline can be estimated as follows. \cite{whi97} concluded that the visual extinction due to diffuse clouds in the direction of Sgr A* is $\sim$20 mag.  As the 3.4 $\mu$m absorption feature arises in diffuse gas, one may assume that diffuse cloud extinction on sightlines toward objects within the CMZ scales with the strength of this feature. Comparison of the depth of the 3.4 $\mu$m feature toward 2M1747 with the depths toward the more central GC sources observed by \cite{san91} and \cite{pen94} and toward the Quintuplet Cluster (Figure 8), all of which have similar values, indicates that the extinction toward 2M1747 due to dust in diffuse clouds is about 25\% greater than toward the other objects, and thus is $\sim$25 mag. 

To determine the total extinction to 2M1747 one must add to the above two values (extinction due to dust in diffuse gas and dust associated with gaseous CO, 25 mag and 160 mag, respectively), the extinction due to dust associated with the solid CO. As discussed in Section 4.2, the presence of solid CO is evident in Figure 4, but the feature itself is partially masked by the strong interstellar CO absorption lines. If the gas-phase lines were not present the solid CO feature, whose central wavelength \citep{lac84,ker93} is nearly coincident with the $1-0$ $P$(1) line at 4.674 $\mu$m, but which usually possesses a long wavelength shoulder due to CO mixed with polar molecules such as H$_2$O, would be much more prominent.

To estimate the column density of solid CO, $N$(CO$_{\rm solid}$) we use the formula $N$(CO$_{\rm solid}$) = $\tau$ $\Delta \nu$ / A , where $\tau$ is the peak optical depth of the solid CO feature, $\Delta \nu$ is the FWHM of its absorption band, and A is the band absorbance per frozen CO molecule. We estimate $\tau$ = 0.3 from Figure 4 and adopt  8 cm$^{-1}$ and $1.1 \times  10^{-17}$ cm$^{-1}$ molecule$^{-1}$, respectively, from \cite{tex99}. Then $N$(CO$_{\rm solid}$) = $2 \times 10^{17}$ cm$^{-2}$, which is less than one percent of the total column density of gaseous CO derived above. The carrier of XCN is probably present in the same grains containing the solid CO. CO may also effectively be locked up in solid CO$_2$, which produces prominent absorptions at 4.27 $\mu$m and 15.2 $\mu$m in the spectrum of the Galactic center \citep{lut96}.  The strengths of the solid CO$_2$ features on the sightline to 2M1747 are not known. However, the  column densities of solid CO$_2$ found by \cite{ger99} toward sources in the Galactic center are somewhat less than that of solid CO. In addition, the ices containing solid CO$_2$ are also probably the same ones that contain solid CO. 

It is thus clear that it is dust associated with the gaseous CO in molecular clouds and dust in the diffuse interstellar medium that produce the vast majority of the extinction. Based on all of the above considerations, we estimate that visual extinction to 2M1747 is 205 mag, with an uncertainty of 75 mag stemming almost entirely from the uncertainty in $N_{gas}$(CO)/ $A_V$ in dense clouds observed by \cite{lac17}.  In addition, as mentioned earlier, the Lacy et al. value may not hold in the Galactic center, where the overwhelming majority of the extinction to 2M1747 arises.

\subsection{Absolute Brightness of 2M1747, the Optical - Infrared Extinction Curve, and a Possible Revised Extinction}

At wavelengths longward of the $H$ band the infrared brightness of 2M1747 is dominated by emission from its circumstellar dust. Nevertheless, one can crudely estimate what the apparent brightness of the unobscured star would be in the $K$ band based on the depth of the $2.3-2.4$ $\mu$m photospheric CO absorption in Figure 3. To do so we assume that the photosphere of 2M1747 is that of a late -ype giant or supergiant, whose CO band head's depths are $\sim$50\%\ of the continuum at a resolving power R$\sim$3,000 \citep{wal97}. That is five times the depth of the veiled CO $2-0$ band head in 2M1747 when smoothed to the same resolving power, which implies that warm dust contributes $\sim$80\%\ of the continuum at 2.3 $\mu$m). This indicates that $K \approx\ 12.2$ mag for an unveiled 2M1747.

To estimate the intrinsic brightness of 2M1747 we use the optical$-$mid-infrared extinction law derived by \cite{wan19}, noting that the infrared portion from 1.0 $\mu$m to 2.5 $\mu$m is similar to that of the near-infrared extinction curves of \cite{fri11} and \cite{nag19}.  For $A_V$ = 205 mag, $A_K$ = 14.8 mag, implying that the unveiled $K$ magnitude of 2M1747 corrected for extinction is $\sim-2.5$. Several OH/IR supergiants, including VY CMa and VX Sgr are among the brightest M supergiants, which have absolute $K$ magnitudes of  $-11$ to $-13$ \citep[e.g][]{cha19},  corresponding to apparent $K$ magnitudes of $1.5 - 3.5$  at the Galactic center, roughly $40-250$ times fainter than an unobscured 2M1747 would be. It seems to us unreasonable that 2M1747 would be so much brighter than them.  Alternatively, if  $A_V$ = 130 mag toward 2M1747 (a deviation of 1$\sigma$ from the estimate in Section 5.1, the unobscured $K$ magnitude of 2M1747 would be $\sim$2.8, comparable to the brightest OH/IR supergiants). 

The above alternatives suggest three possible explanations. If the absolute $K$ magnitude of 2M1747 is comparable to those of the  brightest M supergiants then the visual extinction to 2M1747 is $\sim$130 mag, $\sim$40\%\ lower than the original estimate, and the mean value of $N$(CO$_{\rm gas}$/ $A_V$ derived by \cite{lac17} for the Taurus Molecular Cloud is $\sim$40\%\ lower than the value for the dense gas and dust within the CMZ. We already noted that their value of this ratio is high compared to other recent estimates. Alternatively, the recently derived extinction laws mentioned above, while appropriate for dust in the Galactic plane along the line of sight to the GC, do not apply to the dust in the dense gas within the CMZ, and in particular in Sgr B1. The final possibility is that the visual extinction of 205 mag to 2M1747 and the extinction law are both approximately correct and 2M1747 is considerably more than an order of magnitude more luminous than any other OH/IR star. We regard the third possibility as unlikely, but are unable to conclude whether one of the first two possibilities is more likely than the other.

\subsection{Locations of the Gas and Dust Producing the Solid-state Absorption Bands}

In this subsection we attempt to link several of the well-studied solid-state absorption features observed toward 2M1747 to environments present on its line of sight. Our conclusions regarding their locations are summarized in Table 2. Because of the uncertainties regarding the identifications of the absorption features in the $3.2-3.3$ $\mu$m and $3.8-4.0$ $\mu$m regions we do not speculate on the locations where they arise.  The reader is referred back to Section 4.5 for discussion of the issues regarding their identifications.

Based in part on the spectroscopy reported here and by \citet{geb10} and \cite{oka19} one can identify several environments shown in Figure 1 that potentially contribute to the solid-state absorptions. 
\begin{itemize}
\item Warm dense molecular gas in the Sgr B complex, where 2M1747 is located. Typical gas temperatures outside of the dense cores in Sgr B are $70-200$ K \citep[][this paper]{lis01}. Most of the extinction toward 2M1747 is due to dust in bf this gas.
\item Warm diffuse molecular and atomic gas filling much of the CMZ and thus mostly unassociated with dense molecular clouds there. The mean density of this gas is $\sim$ 50 cm$^{-3}$ and its mean temperature is $\sim$200 K \citep{oka19}
\item Cold and dense molecular gas in the three intervening spiral arms, with $T$ $\sim$ 20 K \citep{ser86}
\item Cold diffuse molecular gas in the three intervening spiral arms, with $T$ $\sim$ 50 K \citep{sno06}
\end{itemize}

As illustrated in Figure 1 dense molecular cores are also present, in Sgr B \citep{bal10} and in the spiral arms, but are not on the line of sight to 2M1747. Low-density interarm gas makes up a majority of the length of the sightline, but the dust it contains probably contributes little to the solid-state absorption features. 

The circumstellar dust associated with 2M1747, which produces most of the continuum in the $K$-band (see Section 5.2) and virtually all of the continuum in the $L$ and $M$ bands, can be easily ruled out as a site in which any of the solid-state featured reported here arise. The characteristic temperature of this dust, estimated from Spitzer photometry in the IRAC1 and IRAC3 filter bands (given in Section 2; we do not use IRAC2, as its wavelength range is coincident with the strong interstellar CO absorption) is $\sim$300 K, far too warm for dust mantles to be present.

\subsubsection{Solid CO and XCN}

Of the solid-state absorption features observed toward 2M1747 and listed in Table 2, two that have been found on grain mantles in many cold dense clouds outside of the GC are the solid CO and XCN absorptions at $4.6-4.7$ $\mu$m. These two bands have been observed not only toward 2M1747, but also in large-aperture spectra of the Central Cluster \citep{mon01} and toward many individual sources within that cluster \citep{mou09}. The strength of solid CO absorption toward 2M1747 is similar to that toward the Central Cluster in the large ISO SWS aperture \citep{mon01}, whereas the XCN band toward 2M1747 is roughly twice as strong as its counterpart. The solid CO and XCN bands observed by \cite{mou09} toward individual sources in and near the Central Cluster have a wide range of strengths, some greater than and some weaker than the bands in the spectrum of 2M1747.

\cite{mon01} concluded that the solid CO absorption toward Sgr A* arises in the foreground spiral arms. This seems reasonable in that the gas temperatures in the spiral arms found by \cite{ser86} are $14-19$ K and that dust temperatures are generally lower than gas temperatures. These temperatures are low enough for CO to adhere strongly to grains. \cite{mon01} also concluded that the absorption by gaseous CO in its fundamental band toward Sgr A* is mostly due to dense gas in the foreground spiral arms.  The high-resolution spectra of optically thin overtone $^{12}$CO and fundamental band $^{13}$CO lines toward the Central Cluster sources GCIRS3 and GCIRS1W \citep{got14, oka19}, which are dominated by absorptions at the characteristic radial velocities of those arms, confirm their conclusion. 

The low-resolution spectra of \cite{mou09} show significant variations in the strengths of the low-$J$ fundamental band lines of $^{12}$CO in addition to variations in the solid CO feature. \cite{mou19} argue that some of the source-to-source variations arise in gas and dust close to the center. This is certainly the case for the gaseous CO, where at high spectral resolution  \cite{got14} observed CO absorption at velocities not observed on other sightlines. They attributed the additional absorption component to gas associated with the 2 pc diameter circumnuclear ring (CNR). The temperature of the gas and dust local to Sgr A* and the Central Cluster is uncertain but it seems unlikely that it would be low enough for solid CO to survive.  Thus, it appears to us to be more likely that the variations in the solid CO band strength toward the sources observed by \cite{mou09} are entirely due to clumpy distributions of solid CO in the three intervening spiral arms. 

Toward 2M1747 the majority of the absorption by gaseous CO comes from the front portion of  Sgr B1, as discussed earlier. From the far-infrared mapping of \cite{lis01}, the mean dust temperature in that portion of Sgr B1 is $\sim$35 K. Because CO rapidly sublimes from dust grains at temperatures above 20 K \citep{san90,dis96}, it is unlikely that the solid CO seen toward 2M1747 arises in gas at that temperature. However, the high extinction and absence of dense cores on this sightline suggests that the length of the absorbing column in Sgr B1 is likely to be several pc and thus one cannot rule out the possibility that there are small shielded regions of lower temperature in one or more parts of the sightline within Sgr B1, where CO ice is present. 

We conclude that it is most probable that the vast majority of the solid CO absorption arises outside of the CMZ, as we and \cite{mon01} conclude is the case for sightlines toward Central Cluster sources.  Because Sgr B1 is 85 pc distant from the Central Cluster sources, apart from gas local to the solar neighborhood the sightline to 2M1747  passes through the intervening spiral arms tens of parsecs distant from the sightlines to the Central Cluster. The most likely location for the solid CO on the 2M1747 sightline would appear to be cold dense molecular gas in the 3 kpc arm, which in the low-$J$ lines of the CO overtone band, absorbing near $-43$ km s$^{-1}$,  produce much deeper absorptions than observed in the other intervening arms (see Figure 2). 

Interstellar XCN is believed to be created by chemical reactions involving CO and/or CH$_3$OH on the mantles of grains in dense molecular clouds. This view is supported by a number of laboratory experiments \citep[e.g.][]{lac84,ber95,sch97,ber00,pal01,hud18}. Methanol ice is not detected on the 2M1747 sightline, which suggests that the production of XCN on this sightline is tied to the presence of solid CO. As discussed earlier, the high mean dust temperature in Sgr B1 on the line of sight to 2M1747 suggests that CO is not present in icy grain mantles there, except possibly for small shielded pockets within the cloud. However, laboratory experiments \cite[e.g.][]{lac84} show that XCN survives at much higher temperatures than solid CO and thus could be prominent in regions where solid CO is not. \cite{chi02} and \cite{mou04} have suggested that the XCN bands observed toward sources in and near the Central Cluster arise at least in part in molecular gas close to Sgr A*, such as the CNR.

If the XCN band observed toward 2M1747 arises in dust mantles in dense clouds outside of the CMZ, the most likely location would appear to be in the 3 kpc arm as argued above for solid CO. However, XCN bands tend to be weak in quiescent cold dark clouds, with strong bands such as that observed toward 2M1747 found in absorption toward massive YSOs embedded in dense clouds and weaker bands toward lower-mass YSOs  \citep{whi01,pon03}. No YSOs are known in those parts of the foreground spiral arms close to the sightline to 2M1747, but we are unaware of any searches for them. In conclusion, we regard the location of the XCN on the sightline toward 2M1747 to be undetermined and possibly present in both the CMZ, within dense regions in Sgr B1 and/or the CNR, and one or more foreground spiral arms. 
 
\subsubsection{H$_2$O}

Toward 2M1747 the broad 3 $\mu$m band of water ice plus impurities in Figure 8 has a peak optical depth of 1.25. This is 2.5 times that observed by ISO SWS toward the Central Cluster \cite{lut96} and $1.5-3$ times the optical depths observed toward individual sources in that  cluster \citep{mcf89,chi02,mou04}, with the exception of the anomalous source IRS 19. This suggests that some of the water ice on the line of sight to 2M1747 is produced in the CMZ. In the Galactic plane the water ice band has not been detected in diffuse molecular gas \cite[e.g][]{whi97} but is present in dense clouds with $A_V > $ 3.3 mag \citep[e.g.][]{whi88}. Thus, the most likely locations for water ice-covered grains  to be found are dense molecular gas in the spiral arms and in Sgr B1. \cite{whi97} estimate that dust in the dense molecular gas on the line of sight to the GC produces $\sim$10 mag of optical extinction, which suggests that the water ice band produced in it could have an optical depth of several tenths.

Given the order-of-magnitude-larger extinction of the molecular gas in Sgr B than the molecular gas in the foreground spiral arms and the mean dust temperature of $\sim$35 K measured in this region of Sgr B1 by \cite{lis01}, it is perhaps surprising that the optical depth of its water ice band is not much greater than the observed value. The lack of very strong absorption might be attributable to an environment inhibiting formation of ice mantles on grains on much of the sightline within Sgr B1. The gas temperature in Galactic center molecular clouds sigificantly higher than in many dense clouds in the Galactic plane. In addition, a cosmic-ray ionization rate of H$_2$ $\sim$10$^3$ higher than in dense clouds in the Galactic plane exists throughout much of the CMZ \citep{pet16,oka19}. Sputtering of water ice off of grain mantles due to cosmic-ray - dust particle collisions \citep{dar15} could be responsible for the lower than expected strength of the 3.0 $\mu$m ice band.

\subsubsection{The 3.4 $\mu$m feature}

In Figure 9 the optical depth spectrum of the 3.4 $\mu$m absorption toward 2M1747 is compared with the mean spectrum of six members of the Quintuplet Cluster observed by \cite{chi13}. The profiles are nearly identical in the $3.35-3.60$ $\mu$m interval, with the absorption toward 2M1747 roughly 30\%\ stronger. \cite{chi02} and \cite{mou04} have presented optical depth spectra of individual infrared sources in the Central Cluster. Both papers reveal considerable source-to-source variation in the strength of the feature. They differ in their interpretations of this. \cite{chi02} ascribe the differences in optical depth to spatial variations the abundance of the carrier of the feature, saturated aliphatic hydrocarbons, in foreground material, while \cite{mou04} argue that not only this absorption feature but also the 3.0 $\mu$m water ice feature arise largely in gas within the Central Cluster itself, within 0.5 pc (12\arcsec) of Sgr A*. The latter is in conflict with the generally accepted view that grains with water ice mantles are found in dense molecular gas and observational evidence that the carrier of the 3.4 $\mu$m feature exists only in diffuse gas. Although the two groups used similar techniques in determining the optical depth of the 3.4 $\mu$m feature, their estimates toward the three infrared sources that both groups observed differ considerably, in one case by almost a factor of three, with the difference much greater than the stated uncertainties. 

The peak optical depth of the 3.4 $\mu$m band toward 2M1747 was determined using the same technique as in the above two papers. Within the uncertainties it  is the same as the average value found by \cite{chi02} for all of their measurements and is about three-fourths of the average value found by \cite{mou04} toward their sources. Because of the roughly equal optical depths we think it is most probable that the majority of the 3.4 $\mu$m absorption toward all of the Galactic center sources arises in the same regions of their sightlines. The sightlines to all of these sources include diffuse gas both within and external to the CMZ, and thus both are candidate sites. However, laboratory studies \citep[][see also references therein]{god11} suggest that the intense cosmic-ray flux and UV field in the CMZ will inhibit formation of the carriers of the 3.4 $\mu$m feature. This suggests that the feature arises in dust within the diffuse gas in the Galactic spiral arms.
 
 \section{Summary}
 
We have presented and analyzed a variety of infrared spectra in selected portions of the $2-5$ $\mu$m wavelength interval toward 2M1747, a luminous OH/IR star located in the CMZ on a sightline to the Sgr B1 giant molecular cloud. The spectra reveal numerous gaseous absorption lines of H$_3^+$ and CO and broad absorption features due to at least seven distinct species associated with interstellar dust grains and their icy mantles.  The gaseous line profiles of both species are dominated by strong absorption at positive radial velocities previously associated with Sgr B1 from radio and millimeter wavelength studies. This implies that 2M1747 is situated within, but toward the front of the dense and warm gas in Sgr B1, whose dust is associated with $\sim$10$^{3}$ visual magnitudes of extinction. At negative velocities, absorption in foreground spiral arms produce sharp absorption components in the lowest-lying lines of both H$_3^+$ and CO. The H$_3^+$ line profiles demonstrate that the sightline to 2M1747 also passes through a long column of warm diffuse gas with properties including kinematics that are consistent with those of the outwardly expanding gas observed previously on many other CMZ sightlines \citep{oka19, oka20}. The total visual extinction to 2M1747 probably exceeds 100 mag.  We propose that the 3 $\mu$m water ice and 4.62 $\mu$m XCN bands observed toward 2M1747 arise both in Sgr B1 and in one or more Galactic foreground spiral arms, but that the solid CO feature arises entirely or almost entirely in the spiral arms. The 3.4  $\mu$m hydrocarbon absorption appears to be produced almost entirely in cold diffuse gas in the arms and not in the warm, irradiated diffuse gas of the CMZ. The observed solid-state features also include a shallow band extending from 3.22 $\mu$m to 3.30 $\mu$m, which may be a blending of two unrelated features, and two weak absorption bands in the $3.8-4.0$ $\mu$m interval, whose previous identifications as CH$_{3}$OH ice now appear to be in doubt as a result of these observations.
 
\acknowledgments

This research is based in large part on observations obtained at the international Gemini Observatory, a program of NSF's NOIRLab, which is managed by the Association of Universities for Research in Astronomy (AURA) under a cooperative agreement with the National Science Foundation, on behalf of the Gemini Observatory partnership: the National Science Foundation (United States), National Research Council (Canada), Agencia Nacional de Investigaci\'{o}n y Desarrollo (Chile), Ministerio de Ciencia, Tecnolog\'{i}a e Innovaci\'{o}n (Argentina), Minist\'{e}rio da Ci\^{e}ncia, Tecnologia, Inova\c{c}\~{o}es e Comunica\c{c}\~{o}es (Brazil), and Korea Astronomy and Space Science Institute (Republic of Korea). We thank the staffs of the Gemini Observatory and the NASA Infrared Telescope Facility for their support. We also thank the graphic artist, Michael Schisler, for the creation of Figure 1. We are grateful to both Takeshi Oka and the referee for helpful comments.

\facilities{Gemini: Gillett; NASA Infrared Telescope Facility}

\software{Figaro  \citep{figaro}, Gemini IRAF Package}

\end{document}